# Systematic definition and classification of data anomalies in DBMS (English Version)

LI Hai-Xiang[1], LI Xiao-Yan[2], LIU Chang[1], DU Xiao-Yong[34], LU Wei[34], Pan An-Qun[1]

[1](Tencent Technology (Beijing) Co., Ltd, Beijing 100080, China)
[2](Department of Information Sciences, School of Mathematical Sciences, Peking University, Beijing 100871, China)
[3](Key Laboratory of Data Engineering and Knowledge Engineering of the Ministry of Education (Renmin University of China), Beijing 100872, China)
[4](School of Information, Renmin University of China, Beijing 100872, China)
Corresponding Author: LI Hai-Xiang, E-mail: blueseali@tencent.com

**Abstract**: There is no unified definition of Data anomalies, which refers to the specific data operation mode that may violate the consistency of the database. Known data anomalies include Dirty Write, Dirty Read, Non-repeatable Read, Phantom, Read Skew and Write Skew, etc. In order to improve the efficiency of concurrency control algorithms, data anomalies are also used to define the isolation levels, because the weak isolation level can improve the efficiency of transaction processing systems. This paper systematically studies the data anomalies and the corresponding isolation levels. We report twenty-two new data anomalies that other papers have not reported, and all data anomalies are classified miraculously. Based on the classification of data anomalies, two new isolation levels system with different granularity are proposed, which reveals the rule of defining isolation levels based on data anomalies and makes the cognition of data anomalies and isolation levels more concise.

## 1 Introduction

The isolation levels of the database are first defined by eliminating specific data anomalies. ANSI-SQL standard [16] describes four data anomalies: *Dirty Write*, *Dirty Read*, *Non-Repeatable Read*, *Phantom Read*, which define four isolation levels: *Read Uncommitted* (RU), *Read Committed* (RC), and *Repeatable Read* (RR), *Serializable*, respectively. For example, the RC isolation level avoids dirty write and dirty read anomalies, while the serializable isolation level avoids all data anomalies. Most commercial database systems support these four isolation levels, such as DB2, Informix, MySQL, PostgreSQL, TDSQL. These database systems, by default, configure a non-serializable isolation level as the weaker isolation level performs better throughput of the database transaction processing.

James Gray et al. [18] pointed out that the ANSI-SQL standard lacks a formal description of data anomalies, leading to some misunderstandings of data anomalies. Therefore, based on four data anomalies by ANSI-SQL, they proposed four new data anomalies: Lost Update, Cursor Lost Update, Read Skew, and Write Skew. To best of our knowledge, we summary 18 publicly reported data anomalies in the-state-of-art papers [3,11,16,18,19,20,21,22,23,25,31,33], as shown in Table 1.

Table 1  Summary of known data anomalies

| NO | Anomaly (Name, Reference, Year) | Formal Definition |
|---|---|---|
| 1 | Dirty Write,[16] 1992,[18] 1995 | $W_1[x] \dots W_2[x] \dots (C_1 \text{ or } A_1) \text{ and } (C_2 \text{ or } A_2) \text{ in any order}$ |
| 2 | Lost Update,[18] 1995 | $R_1[x] \dots W_2[x] \dots W_1[x] \dots C_1$ |
| 3 | Dirty Read,[16] 1992,[18] 1995 | $W_1[x] \dots R_2[x] \dots (A_1 \text{ and } C_2 \text{ in either order})$ |
| 4 | Aborted Reads,[20] 2000,[19] 2015 | $W_1(x1:i) \dots R_2(x1:i) \dots (A_1 \text{ and } C_2 \text{ in any order})$ |
| 5 | Fuzzy OR Non-Repeatable Read,[16] 1992 | $R_1[x] \dots W_2[x] \dots C_2 \dots R_1[x] \dots C_1$ |
| 6 | Phantom,[16] 1992 | $R_1[P] \dots P_2[y \text{ in } P] \dots C_2 \dots R_1[P] \dots C_1$ |
| 7 | Intermediate Reads,[20] 2000,[19] 2015 | $W_1(x1:i) \dots R_2(x1:i) \dots W_1(x1:j) \dots C_2$ |
| 8 | Read Skew,[18] 1995 | $R_1[x] \dots W_2[x] \dots W_2[y] \dots C_2 \dots R_1[y] \dots (C_1 \text{ or } A_1)$ |
| 9 | Unamed Anomaly,[21] 2000 | $R_3[y]R_1[x]W_1[x]R_1[y]W_1[y]C_1 \ R_2[x]W_2[x]R_2[z]W_2[z]C_2 \ R_3[z]C_3$ |
| 10 | fractured reads,[31] 2014,[3] 2017 | $R_1[x_0] \dots W_2[x_1] \dots W_2[y_1] \dots C_2 \dots R_1[y_1]$ |
| 11 | Serial-Concurrent-Phenomenon,[25] 2014 | $R_1[x_0] \dots W_2[x_1] \dots W_2[y_1] \dots C_2 \dots R_1[y_1]$ |
| 12 | Cross-Phenomenon,[25] 2014 | $R_1[x_0] \dots R_2[y_0] \dots W_3[x_1] \dots C_3 \dots W_4[y_1]C_4 \dots R_2[x_1] \dots R_1[y_1]$ |
| 13 | long fork anomaly,[3] 2017 | $R_4[x_0] \dots W_1[x_1] \dots R_3[y_0] \dots R_3[x_1] \dots W_2[y_1] \dots R_4[y_1]$ |
| 14 | causality violation anomaly,[3] 2017 | $R_3[x_0] \dots W_1[x_1] \dots C_1 \dots R_2[x_1] \dots W_2[y_1] \dots C_2 \dots R_3[y_1]$ |
| 15 | A read-only transaction anomaly,[11] 1982,[22] 2004 | $R_2(x_0,0)R_2(y_0,0)R_1(y_0,0)W_1(y_1,20)C_1 \ R_3(x_0,0)R_3(y_1,20)C_3 \ W_2(x_2,-11)C_2$ |
| 16 | Write Skew,[18] 1995 | $R_1[x] \dots R_2[y] \dots W_1[y] \dots W_2[x] \dots (C_1 \text{ and } C_2 \text{ occur})$ |
| 17 | Predicate-Based Write Skew,[23] 2005 | $R_1(\{x_0 \text{ in } P\}) \dots R_2(\{y_0 \text{ in } P\}) \dots W_1[\{y_1 \text{ in } P\}] \dots C_1 \dots W_2[\{x_1 \text{ in } P\}]$ |
| 18 | Read partial committed anomaly ,[33] 2019 | $R_1[x] \dots W_2[x] \dots W_2[y] \dots C_2 \dots R_1[y] \dots C_1$ |

Note that this paper is an English version of published paper [34]. This paper contributes to systematically study the data anomalies. We define the data anomalies with a unified method as well as propose 22 unreported data anomalies. We study the characteristics of data anomalies, and classify them. The classification methods based on data anomalies demonstrate how to define the isolation level in different granularity based on data anomalies. Section 2 describes the related work. Section 3 gives the framework for describing data anomalies, and formalizes the definition of data anomalies, and classifies them. Section 4 defines the isolation level based on the classification of data anomalies. Section 5 summarizes the paper and gives acknowledgment.

## 2  Related Work

Previous studies have not systematically studied data anomalies. Mostly, they research data anomalies from the following perspectives.

**New data anomalies are often reported in the form of cases, instead of systematic research.** Data anomalies have not been systematically studied. The understanding of data anomalies remains in the form of cases. As shown in Table 1, new data anomalies were reported in 2000, 2005, 2014, and 2017, indicating that the research on data anomalies has been never-ending.

The four data anomalies proposed in the ANSI-SQL standard [16] are caused by a single variable. Anomalies such as *Intermediate Reads* and *Abort Read* proposed in [19,20] are also single-variable anomalies. However, anomalies such as *Read Skew* and *Write skew* proposed in [18] are caused by two variables. [21] even proposed a three-variable anomaly, which is relevant to the read skew anomaly. [26] discussed more variants of write skew anomaly in PostgreSQL. [11,22] describes a read-only transaction anomaly caused by two variables and three concurrent transactions. [11,22,23,25,3,31] proposes a two-variable anomaly with three or four transactions, yet without further study whether more variables will cause new data anomalies. Although data anomalies seem to be related to the number of variables, the relationship between the number of variables and anomalies remains unknown.

Two distributed data anomalies, *Serial Concurrent Phenomenon* and *Cross Phenomenon*, are first proposed in [25]. Later, similar data anomalies, *Long Fork* and *Causality Violation*, are proposed in [3]. *Fractured Reads* anomaly, which is similar to read skew is proposed in [31]. *Predicate-Based Write Skew* anomaly based on two variables and violating constraints are described in [23]. The booming discovery of these data anomalies drives us to several questions. What is the root cause of distributed data anomalies? What are the differences and similarities between distributed and non-distributed data anomalies? Why do researchers give different names to seemingly similar data anomalies? Furthermore, what is the ultimate number of data anomalies and what is the essential relation between these anomalies? Hopefully, we can give you the solutions or ideas to these questions via this paper.

**Data anomaly and serializability theory.** A formal definition of serializability is proposed in [1]. Concurrent transactions guarantee no data anomalies only if they meet the serializable scheduling. To the best of our knowledge, there exists only one paper [4] describing the definition of data anomalies by using the concept of serialization. For the solution of serializability, [5] pointed out that the solution by direct serialization graphs (DSG) is an NP problem. The concept of serializability has the theoretical meaning of verifying whether a giving scheduling guarantees the data consistency [2,32], yet it is impractical in production. Later, the conflict serialization is discussed in [15,30]. The conflict serializability is a sufficient but not necessary condition for serializability. According to the serialization graph [15], precedence graph [24], dependency graph [18], and DSG [20,23], conflict serializability has practical meaning to guide the implementation of the transaction scheduling engine. However, although the idea of conflict serializability is often integrated into 2PL [1], T/O [12], and other concurrent control algorithms to guarantee serializability, it leads to the increase of false rollback. In this paper, the concepts of conflict relationship and conflict serializable graph are extended. The *partial order pair* and *partial order pair cycle graph* are formally introduced later. These partial order pairs and partial order pair cyclic graphs help to formally define data anomalies.

**Data anomalies and isolation levels.** Isolation levels are often defined by known data anomalies [16,18,20]. However, we later show countless data anomalies exist via the partial order cyclic graph constructed by partial order pairs. The isolation levels usually do not cover all these known anomalies, not to mention those unknown ones. In this paper, by formally describing data anomalies, we give a formal definition of all data anomalies. By a unified way of definition and classification, we then can holistically and concisely define the isolation levels towards all data anomalies.

**Quantification and classification of data anomalies.** The existing research work quantifies and classifies data anomalies based on the known and limited numbers of data anomalies (only including some partial data anomalies shown in Table 1) by using serializability theory [6] and dependency graph. [12] developed a set of tools and methods, which automatically detects data anomalies under snapshot isolation. There also proposed embedding middleware between applications and databases [8,9,10,13], to quantify and classify data anomalies in specific scenarios such as TPC-C. These solutions do not and cannot find new data anomalies. [7] quantitatively studies the rate of integrity violation of data anomalies under snapshot isolation at different isolation levels. Again, it does not find new data anomalies. Based on DSG and some data anomaly cases, [20] defines some phenomena related to data anomalies. However, there is no further systematic study of data anomalies. In this paper, we can define and classify all data anomalies. Also, based on our definition and classification, we will explain the relationship between data anomalies, isolation levels, and concurrent control algorithms.

**This paper aims to systematically study data anomalies.** In the following, we systematically study the characteristic of data anomalies, define data anomalies in a unified way. We will classify all data anomalies, show 22 new (never-reported) data anomalies, and define isolation levels towards all these classified data anomalies.

## 3  Formal Definition and Classification of Data Anomalies

The concurrent control mechanism is to schedule concurrent operations in a way that the execution of a user's transaction is not disturbed by other transactions, avoiding data inconsistency, i.e., data anomalies. Before we begin to define data anomalies, we first express some symbols that will be used throughout the paper. Let $D = \{x, y, z ...\}$ denote the variables, which are the data objects in databases. A variable can be a tuple/record, a page, or even a table object. After a variable is written, its state will change, calling a new version of the variable. Different versions of a variable are represented by subscripts, i.e., $x_0, x_1, x_2$. Let $T = \{t_1, t_2, ..., t_n\}$ denote the set of transactions, and for each transaction $t_i = (op_i, <_s)$, where $op_i$ is the set of operations and states in $t_i$. For example, $p_i <_s q_j$ meaning operation $p_i$ precedes $q_j$ in schedule $s$.

### 3.1  Definition

**Definition1: Operation**

The operations of a transaction on a variable consist of read (R) and write (W). Let $R_i[x_m]$ and $W_j[y_n]$ denote the read operation on variable $x$ in transaction $t_i$ and the write operation on the variable $y$ in the transaction $t_j$, respectively, where the subscript of the read/write operation is exactly the transaction number it belongs to. The states of transactions consist of Commit (C), Abort (A), and Undone (U). Let $C_i$, $A_i$, and $U_i$ denote the committed, aborted, and undone status of $t_i$. Usually, the states of undone will not be specified, and there exist subsequent operations in the undone transaction $t_i$.

A commit status indicates that all reads and writes in the transaction are legally operated, meaning the confirmation of a legal status.

An abort status indicates that the transaction needs to discard all reads/writes and rollback to the status it started with, meaning all the written new versions will restore to their old versions.

When there is no need to distinguish $R, W, C, A$ and $U$, $p$ or $q$ denotes an operation or a state, i.e., $p_j = \{R_j[x_n], W_j[x_n], C_j, A_j | x \in D, n \in N, t_j \in T\}$, where $T$ is the set of transactions, $D$ is the set of the operations, and $n$ denotes the version number of the variable. Let $p_j[x_m]$ denotes an operation on $m$'s version of variable $x$ by transaction $t_j$. For short, $p_j$ represents an operation or a status in transaction $t_j$.

**Definition 2: Schedule(s)**

A schedule($s$) refers to a sequence composed of multiple operations or states from a group of transactions.

In a schedule, let $T(s)$, $D(s)$, and $Op(s)$ denote the set of all transactions, the set of all involved variables, and the set of all operations, respectively.

### 3.2  Data anomalies

**Definition 3: Conflicts and Conflicts Relation**

Given any schedule $s$, and two transactions $t_i, t_j \in T(s), t_i \neq t_j$ without any versioning, if exist two operations $p \in t_i$ and $q \in t_j$ operating on the same variable, and at least one operation of $p, q$ is a write, i.e., $(p = R[x] \land q = W[x]) \lor (p = W[x] \land q = R[x]) \lor (p = W[x] \land q = W[x])$. We say $p, q$ is a *conflict* in $s$. The set of all conflicts in $s$ is called *conflict relation*, as defined follows:

$$conf(s) \coloneqq \{(p, q) | p, q \text{ is the conflict in } s, \text{and } p <_s q\}.$$

**Example 1:** The conflict relation of schedule $s = W_1[x_0]W_1[x_1]W_2[x_3]R_1[x_3]C_1C_2$ is $conf(s) = \{(W_1W_2[x]), (W_2R_1[x])\}$.

In the traditional conflict relation, only "read" and "write" operations are considered. However in practice, the conflict relation is also closely related to the "commit" and "abort" of transactions. Therefore, we will add these two operations to extend the definition of conflict relation in scheduling $s$.

**Definition 4: Conflicts and Conflicts Relation with status**

Given any schedule $s$, and $t_i, t_j \in T(s), t_i \neq t_j$, if exists $p \in t_i$ and $q \in t_j$, and p, q is a conflict, denoting $(p_i, q_j)$. Then conflicts can be divided into the following seven categories:

(1)  $p_i - C_i - q_j - U_j/A_j/C_j$: at least one write exists in $p, q$ and $p_i <_s C_i <_s q_j <_s U_j/A_j/C_j$
(2)  $p_i - A_i - q_j - U_j/A_j/C_j$: at least one write exists in $p, q$ and $p_i <_s A_i <_s q_j <_s U_j/A_j/C_j$
(3)  $p_i - q_j - C_i - U_j/A_j/C_j$: at least one write exists in $p, q$ and $p_i <_s q_j <_s C_i <_s U_j/A_j/C_j$
(4)  $p_i - q_j - A_i - U_j/A_j/C_j$: at least one write exists in $p, q$ and $p_i <_s q_j <_s A_i <_s U_j/A_j/C_j$
(5)  $p_i - q_j - C_j - U_i/A_i/C_i$: at least one write exists in $p, q$ and $p_i <_s q_j <_s C_j <_s U_i/A_i/C_i$

(6) $p_i - q_j - A_j - U_i/A_i/C_i$: at least one write exists in $p, q$ and $p_i <_s q_j <_s A_j <_s U_i/A_i/C_i$

(7) $p_i - q_j = \{p_i - U_i - q_j - U_j/A_j/C_j \vee p_i - q_j - U_i - U_j/A_j/C_j \vee p_i - q_j - U_j - U_i/A_i/C_i\}$: at least one write exists in $p, q$ and the transactions they belong to are undone

In (1)-(4), $t_i$ commits or aborts before $t_j$, i.e., $p_i <_s q_j$, denoting $(p_i, o_i, q_j)$ or $(p_i, q_j, o_i), o \in \{A, C\}$. In (5)-(6), $t_j$ commits or aborts before $t_i$, i.e., $p_i <_s q_j$, denoting $(p_i, q_j, o_j), o \in \{A, C\}$. In (7), the transactions in schedule $s$ are undone,, i.e., $p_i <_s q_j$, denoting $(p_i, q_j)$. The above are the conflict categories containing transaction status in $s$. The set all conflicts in such a scheduling $s$ is called the conflict relationship with status, denoting $conf_{ac}(s)$, expressing following:

$conf_{ac}(s) := \{(p_i, o_i, q_j) \vee (p_i, q_j, o_i) \vee (p_i, q_j, o_j) \vee (p_i, q_j) | ((p_i, q_j) \in conf(s)) \wedge (o \in \{A, C\}\}$.

**Definition 5: Conflict Equivalence with Transaction Status**

The two schedulers $s$ and $s'$ are called the conflict equivalence, i.e., $s \approx_{conf_{ac}} s'$, if the two schedulers have the same operations, and the same conflict relations with status. In other words, if $s \approx_{conf_{ac}} s'$, then it needs to meet the following conditions:

(1) $op(s) = op(s')$
(2) $conf_{ac}(s) = conf_{ac}(s')$.

**Definition 6: Conflict Graph with Transaction Status**

We say a graph $G(s) = (V, E)$ is a conflict graph with transaction status in schedule $s$, if it meets:

(1) $V = T(s)$,
(2) $(p_i q_j) \in E \Leftrightarrow t_i \neq t_j \wedge (p_i, q_j) \in conf_{ac}(s)$

**Example 2:** The conflict relation of $s = R_1[x_0]R_1[y_0]W_2[y_1]W_3[z_1]W_1[z_2]C_2W_3[y_2]$ is $conf_{ac}(s) = \{(R_1W_2[y]), (W_3W_1[z]), (R_1W_3[y]), (W_2C_2W_3[y])\}$. Then we show the conflict graph with transaction status in Figure 1.

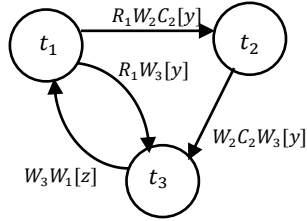

Fig. 1    Conflict graph with transactions status of schedule $s$.

**Definition 7: Partial Order Pair (POP)**

In the case of (2) in the conflict relation with transaction status, due to the in-time abort in $t_i$, $p_i$ has no effect on $q_j$. Likewise in case (6). In other case, if we substitute $(p, q)$ by $\{WW, WR, RW\}$, we can obtain 15 cases of conflicts. We classify these 15 cases into 12 categories, as follows:

(1) $W_i C_i R_j = \{W_i C_i R_j\}$: $t_i$ writes a valid version $x_i$, and $t_j$ reads it;
(2) $W_i C_i W_j = \{W_i C_i W_j\}$: $t_i$ writes a valid version $x_i$ and commit, then $t_j$ writes another valid version $x_j$;
(3) $R_i C_i W_j = \{R_i C_i W_j\}$: $t_i$ reads a version $x_i$ and commit, $t_j$ writes a valid version $x_j$, where there exists a write operation in $t_i$;
(4) $W_i W_j = \{W_i W_j, W_i W_j C_j\}$: $t_i$ writes a version $x_i$, $t_j$ overwrites with another version $x_j$, such that, there may be inconsistence between them;
(5) $W_i R_j = \{W_i R_j, W_i R_j C_j\}$: $t_i$ writes a version $x_i$, then $t_j$ reads it;
(6) $R_i W_j = \{R_i W_j, R_i W_j C_j\}$: $t_i$ reads a version $x_i$, then $t_j$ writes a new version $x_j$, which may affect the read or write of $t_i$ in the same variable;
(7) $W_i R_j A_i = \{W_i R_j A_i\}$: $t_j$ reads a version written by $t_i$, then $t_i$ abort, meaning $t_j$ reads an unexciting version;
(8) $W_i W_j C_i = \{W_i W_j C_i\}$: $t_i$ writes a version $x_i$, and $t_j$ writes another version $x_j$, then $t_i$ commit and update the version by such $x_i$ such that $t_j$ cannot read its version anymore.
(9) $W_i W_j A_i = \{W_i W_j A_i\}$: $t_j$'s written version is overwritten by the old version of $t_i$ due to the abort of $t_i$;
(10) $R_i W_j C_i = \{R_i W_j C_i\}$: $t_i$ reads a version $x_i$, $t_j$ writes a new version $x_j$, $t_i$ commits with the new written version;
(11) $W_i R_j C_i = \{W_i R_j C_i\}$: $t_i$ writes a version $x_i$, $t_j$ reads it and then $t_i$ commits;
(12) $R_i W_j A_i = \{R_i W_j A_i\}$: $t_i$ reads a version $x_i$, $t_j$ writes a new version $x_j$, $t_i$ aborts with the new written version;

From above, each of the cases in (7)-(12) is a cycle. Specifically in (10), the commit of $t_i$ does not effect the write of $t_j$ while it yields POP which is equivalent to (6) $R_iW_j$. Likewise in (11) and (12), they yield POPs which are equivalent to (5) $W_iR_j$ and (6) $R_iW_j$, respectively.

Another cases in (7)-(9), transaction $t_j$ does ghost operation (More detail for anomalies $W_iW_jC_i$ and $W_iW_jA_i$ in Appendix 1). Without losing generality, we detail these three cases in the following.

First in (7), we can divide $\{W_iR_jA_i\}$ into $\{W_iR_j[x]\}$ and $\{R_jA_i[x]\}$, i.e., it is the combination of case (5) and $\{R_jA_i[x]\}$ on the same variable $x$. So we use $R_jA_i$ to denote case (7) for short, as we show its graphic representation in Figure 2.

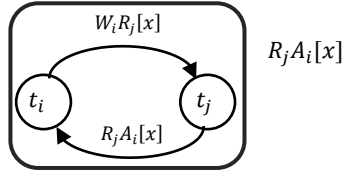

**Fig. 2**  The partial order pair $R_jA_i$ constitutes a partial order cycle.

Then in (8), we can divide $\{W_iW_jC_i[x]\}$ into $\{W_iW_j[x]\}$ and $\{W_jC_i[x]\}$, i.e., it is the combination of case (4) and $\{W_jC_i[x]\}$. We use $W_jC_i$ to denote case (8) for show, as we show its graphic representation in Figure 3.

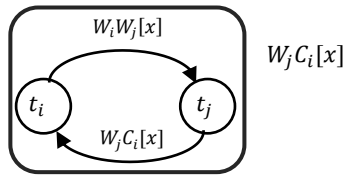

**Fig. 3**  The partial order pair $W_jC_i$ constitutes a partial order cycle.

Finally in (9), we can divide $\{W_iW_jA_i\}$ into $\{W_iW_j[x]\}$ and $\{W_jA_i[x]\}$, i.e., it is the combination of POP (4) and $\{W_jA_i[x]\}$. We use $W_jA_i$ to denote case (9) for short, as we show its graphic representation in Figure 3.

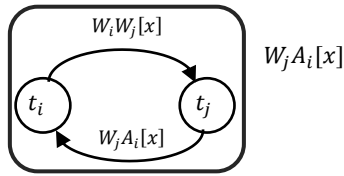

**Fig. 4**  The partial order pair $W_jA_i$ constitutes a partial order cycle.

We name the combinational operations in (1)-(9) the Partial Order Pair (POP). Formally, for a given schedule $s$, let $p_{ij}(s, x)$ or $p_{ij}[x]$ denote POP, where $\forall x \in D(s), \forall p_i <_s q_j,$

We then can let $Pop(s) = \{p_{ij}(s,x) | x \in D(s) \land t_i <_s t_j\}$ or $Pop = \{p(s,x) | s \text{ is an arbitrary schedule} \land x \in D(s)\}$ denote the set of all POPs in schedule $s$.

**Definition 8: Partial Order Pair Graph (PG)**

We say a graph $PG = (V, E)$ is a Partial Order Pair graph in schedule $s$, if it meets:

(1) $V = T(s)$;

(2) $(t_i, t_j) \in E \Leftrightarrow t_i \neq t_j \land \forall x \in D,$
$$p_{ij}[x] \in \{W_iC_iR_j, W_iC_iW_j, R_iC_iW_j, W_iW_j, W_iR_j, R_iW_j, W_iR_jA_i, W_iW_jC_i, W_iW_jA_i\}$$

**Example 3:** Given a schedule $s = W_1[x]R_2[x]A_1W_2[y]C_2R_3[y]$, we can obtain its set of POPs:
$$Pop(s) = \{(W_1R_2A_1[x]), (W_2C_2R_3[y])\}$$

From the set of POPs, we can build the PG in Figure 5

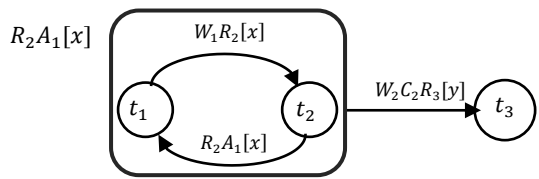

**Fig.5**  Partial order pair graph of schedule $s$.

**Definition 9: Equivalence of POP**

Given schedule $s$ and $s'$, we say they are equivalence, i.e., $s \approx_{PG} s'$, if they have the same operation and POPs. Formally, $s \approx_{PG} s'$, then

(1) $op(s) = op(s')$,
(2) $E(s) = E(s')$.

**Definition 10: Data Anomaly (DA)**

If schedule $s$ has data anomalies, and another schedule $s'$ exists a cycle (including the cycles in POPs), and $s \approx_{PG} s'$.

If exists cycles in PG, then we call it **Partial order pair circle graph(OCG)**, i.e., $PCG(s) = (V, E)$.

According to the definition of PCG, there are at least two transactions in a PCG. Since they construct a directed cycle graph, so they have at least two edges. Apparently, there can be infinite numbers of $T(s)$ in a schedule $s$, making infinite numbers of vertices and edges in PCG. Due to the equivalence between PCG and data anomaly, we come up with the conclusion that there exist infinite data anomalies. Classifying these infinite data anomalies become meaningful and useful for isolation levels. As we can classify the PCG by their edge categories, we can classify data anomalies accordingly.

In practice, we consider PCGs with the least partial order pairs as the anomaly cycle. If many cycles have the same number of pairs, then we consider the first-come operation one as the anomaly cycle.

**Example 4:** The set of POP of schedule $s = R_1[x_0]W_2[x_1]W_2[y_1]W_3[y_2]W_3[z_1]R_1[z_1]R_3[x_1]W_4[x_2]$ is:

$$Pop(s) = \{(R_1W_2[x]), (R_1W_4[x]), (W_2R_3[x]), (W_2W_4[x]), (W_2W_3[y]), (W_3R_1[z])\}.$$

We show the PCG of $s$ in Figure 6.

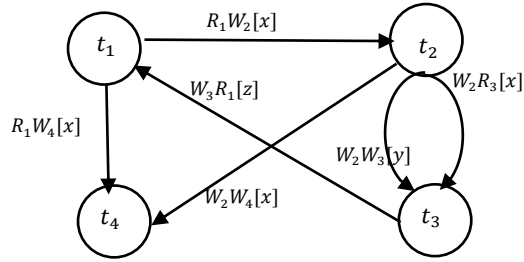

**Fig.6** Partial order pair circle graph of schedule $s_1$.

From schedule $s$, we obtain the cycle graph with least partial order pairs as well as with first-come transactions, i.e., $G = (\{t_1, t_2, t_3\}, \{(R_1W_2[x]), (W_2R_3[x]), (W_3R_1[z])\})$.

**Theorem 10.1: if a PCG is with three transactions with a single variable, then there exists a PCG with two transactions.**

Poof:

We first assume that the PCG with three transactions with a single variable is $G = (\{t_1, t_2, t_3\}, \{(p_1q_2), (p_2q_3), (p_3q_1)\})$. If the POP exists $\{RA, WA, WC\}$, then itself is a 2-transaction PCG.

Since it must have a write operation in partial order pairs and the operation order in schedule is not fixed, so we discuss the following two cases:

(1) When $p_1 = W_1$, it will be a partial order pair no matter what operation $p_2$ is. If $p_1$ precedes $p_2$, since we have partial order $(p_2q_3)$, then we have $p_1 <_s p_2 <_s q_3$. Partial order pair $(p_1q_3)$ and $(p_3q_1)$ construct a 2-transcation cycle. If $p_2$ precedes $p_1$, then Partial order pair $(p_1q_2)$ and $(p_2q_1)$ construct a 2-transaction cycle.

(2) When $p_1 = R_1$, as it exists a partial order pair $(p_1q_2)$, then $q_2 = W_2$. Likewise in (1), If $p_2$ precedes $p_3$, then we have $p_2 <_s p_3 <_s q_1$. Partial order pair $(p_2q_1)$ and $(p_1q_2)$ construct a 2-transaction cycle. If $p_3$ precedes $p_2$, then Partial order pair $(p_2q_3)$ and $(p_3q_2)$ construct a 2-transaction cycle.

So, we conclude that, in any case, there exists a 2-transaction cycle.

Example 5: We show an example in Figure 7 to illustrate the reduction from a 3-transaction PCG to a 2-transaction PCG.

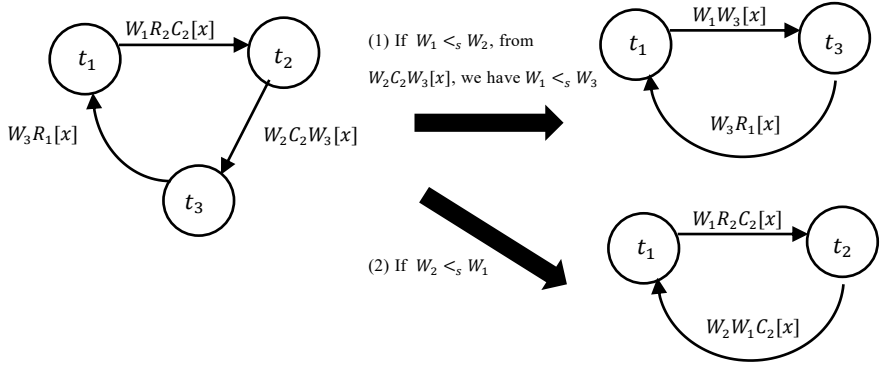

**Fig.7** The single variable partial order by 3 transactions cycle is reduced to the cycle with 2 transactions.

**Theorem 10.2: The single-variable multi($\geq 3$)-transaction partial order pair cycle can reduce to a 2-transaction partial order pair cycle.**

Poof:

We first assume single-variable multi-transaction PCG is $G = (V, E)$, where $V = \{t_1, t_2, \ldots, t_n\}, n \geq 4, V \in T(s)$ and $E = \{(p_1q_2), (p_2q_3), (p_3q_4), \ldots, (p_{n-1}q_n), (p_nq_1)\}$. We then following prove the theorem by induction.

By Theorem 10.1, we already proved that $N_t = 3$ is correct.

We first assume the theorem is correct when $N_t < n$, i.e., $E = (p_1q_2), (p_2q_3), (p_3q_4), \ldots, (p_{k-1}q_k), (p_kq_1), k < n$ can reduce to a 2-transaction cycle $(p_sq_k)(p_kq_s)$.

Now, when $N_t = n$, $E(n) = (p_1q_2), (p_2q_3), (p_3q_4), \ldots, (p_{n-1}q_n), (p_nq_1)$,

(1) When $p_1 = W_1$, it will be a partial order pair no matter what operation $p_{n-1}$ is. If $p_1$ precedes $p_{n-1}$, since we have partial order pair $(p_1q_{n-1})$, then we have $p_1 <_s p_{n-1} <_s q_n$. Partial order pair $(p_1q_n)$ and $(p_nq_1)$ construct a 2-transcation cycle. If $p_{n-1}$ precedes $p_1$, then Partial order pair $(p_{n-1}q_1)$, $(p_1q_2)$ and $(p_{n-1}q_2)$ can reduce to a 2-transaction cycle.

(2) When $p_1 = R_1$, as it exists a partial order pair $(p_1q_2)$, then $q_2 = W_2$. Likewise in (1), If $p_2$ precedes $p_n$, then we have $p_2 <_s p_n <_s q_1$. Partial order pair $(p_2q_1)$ and $(p_1q_2)$ construct a 2-transaction cycle. If $p_n$ precedes $p_2$, we can remove $p_1$, and $E = (p_2q_3), (p_3q_4), \ldots, (p_nq_2)$. By assumption, $N_t = n - 1$ can be reduced to a 2-transcation cycle.

So, we conclude that, in any case, we can reduce to a 2-transaction cycle.

**Example 6:** Figure 8 depicts an example of reduction from 5-transaction PCG to 2-transaction PCG.

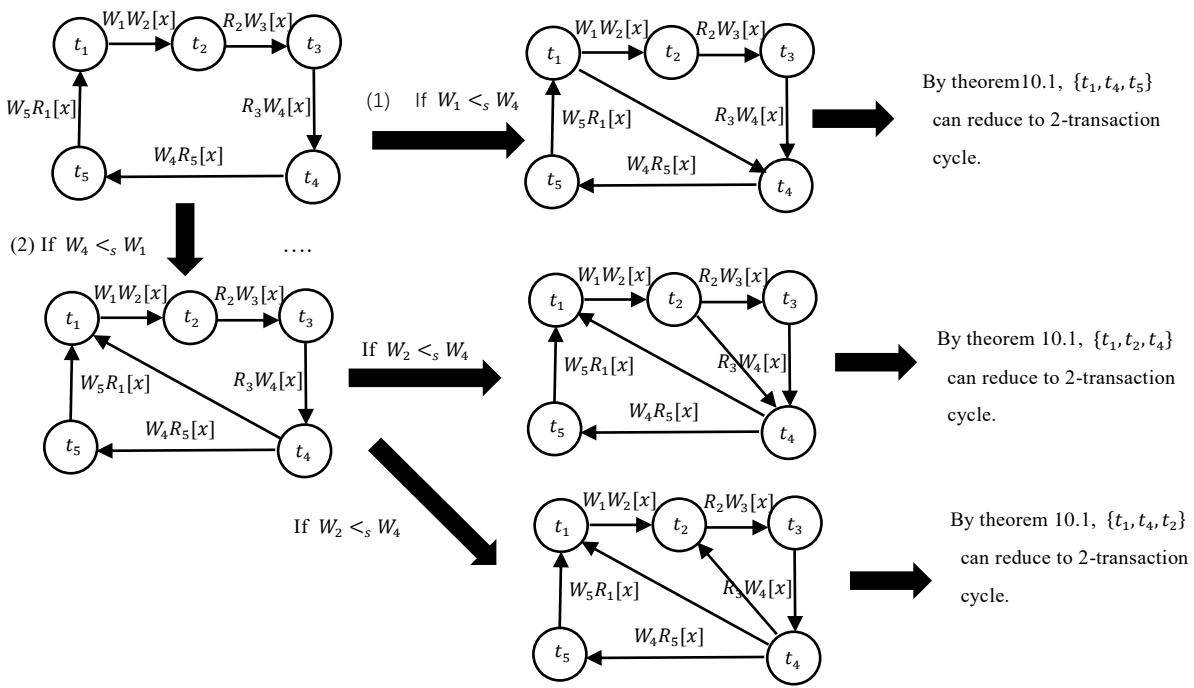

**Fig.8** The single variable partial order by 5-transaction PCG is reduced to 2-transaction PCG.

We show all the symbols and their descriptions in Table 2.

Table 2 Symbols and descriptions.

| Symbol | Description |
|---|---|
| $s$ | schedule |
| $p_i <_s q_j$ | $p_i$ precedes $q_j$ in s |
| $op_i(s)$ | Operation set of $t_i$ in s |
| $p_{ij}[x] = (p_i p_j[x])$ | Partial order pair set by $t_i$ and $t_j$ on variable $x$ in s |
| $W_i[x_m]$ | $t_i$ writes m-th version of $x$ |
| $R_i[x_n]$ | $t_i$ reads n-th version of $x$ |
| $T(s)$ | Transaction set in $s$ |
| $Pop(s)$ | Partial order pair set in $s$ |
| $D(s)$ | Variable set in s |
| $Pop$ | All partial order pair set |
| $PCG$ | Partial order pair cycle graph |
| $N_D$ | Variable count in s |
| $N_T$ | Transaction count in s |
| $C_{da} = (N_D, N_T)$ | Multi-transaction anomaly |
| WAT | Write anomaly type |
| RAT | Read anomaly type |
| IAT | Intersect anomaly type |
| SDA | Single-variable by Double-transactions Anomaly |
| DDA | Double-variable by Double-transactions Anomaly |
| MDA | Multi-variable by Double-transactions Anomaly |

**3.3 Classification of Data Anomalies**

Generally, as the number of transactions, variables, or operations in a schedule increases, an infinite number of data anomalies may occur. Understanding the relation and similarity among these data anomalies helps to comprehend the diversity and internal regularity of data anomalies, leading to a clear classification system. In other words, we can recognize and classify any given data anomalies as well as correlate these anomalies. Also, the classification helps to define isolation levels between these infinite anomalies (details in Section 4). Hopefully, the new isolation levels can re-design the concurrent control algorithms. Based on our new definition, we are able to classify all anomalies and relate them to new isolation levels. Isolation levels can be defined based on different forms in different situations. A simple definition of isolation levels may simplify the design of concurrent control algorithms while more granularity isolation levels help to in-depth grasp at the subtle difference between different data anomaly categories. In fact, giving isolation levels is one of the ways to classify data anomalies.

According to Definition 10, the data anomalies and PCGs are equivalent. So, the classification of PCGs is in fact the classification of data anomalies. We show how we classify these cycles in the following.

**3.3.1 Classification of Single-variable Double-transaction Data Anomalies**

According to theorem 10.1 and theorem 10.2, single-variable PCGs will eventually be reduced to 2-transaction PCG. Therefore, we first classify the 2-transaction PCGs. In fact, when the set Pop(s) contained $\{W_iW_jA_i, W_iW_jC_i, R_iW_jA_i\}$, they are themselves PCGs. So, we classify these three POPs to 2-transaction cycles (details of anomalies (1) and (2) in Appendix 2). In other cases, the 2-transaction cycles $PCG(s) = (\{t_i, t_j\}, \{p_{ij}[x], p_{ji}[x]\})$ can be divided into 9 data anomaly forms (details of anomalies (3)-(11) in Appendix 2), as summaries in Table 3.

Table. 3 PCGs with a single variable and two edges.

| No | Anomaly Name | Formal expression | PG |
|---|---|---|---|
| (1) | Dirty Write | $W_i[x_m] \dots W_j[x_{m+1}] \dots A_i/C_i$ | $W_iW_j - W_jA_i$ |
| | | | $W_iW_j - W_jC_i$ |
| | | | $W_iW_jC_i - R_jW_iC_i$ |
| | | | $W_iW_jC_i - R_jW_iC_i$ |
| | | | $W_iW_jC_i - W_jW_iC_i$ |
| | | | $W_iW_jC_j - R_jW_iC_j$ |
| (2) | Dirty Read | $W_i[x_m] \dots R_j[x_m] \dots A_i$ | $W_iR_j - R_jA_i$ |
| | | | $W_iW_jA_j - R_jW_iA_j$ |
| | | | $R_iW_jA_j - R_jW_iA_j$ |
| (3) | Lost Self Update Committed | $W_i[x_m] \dots W_j[x_{m+1}] \dots C_j \dots R_i[x_{m+1}]$ | $W_iW_jC_j - W_jC_jR_i$ |
| | | | $W_iR_jC_j - W_jC_jR_i$ |
| (4) | Full-Write Committed | $W_i[x_m] \dots W_j[x_{m+1}] \dots C_j \dots W_i[x_{m+2}]$ | $W_iW_jC_j - W_jC_jW_i$ <br> $W_iR_jC_j - W_jC_jW_i$ <br> $W_iW_jC_j - R_jC_jW_i$ <br> $W_iW_jC_j - R_jC_jW_i$ |
| (5) | Non-repeatable Read Committed | $R_i[x_m] \dots W_j[x_{m+1}] \dots C_j \dots R_i[x_{m+1}]$ | $R_iW_jC_j - W_jC_jR_i$ |
| (6) | Lost Update Committed | $R_i[x_m] \dots W_j[x_{m+1}] \dots C_j \dots W_i[x_{m+2}]$ | $R_iW_jC_j - W_jC_jW_i$ <br> $R_iW_jC_j - R_jC_jW_i$ <br> $R_iW_jC_j - R_jC_jW_i$ |
| (7) | Full Write | $W_i[x_m] \dots W_j[x_{m+1}] \dots W_i[x_{m+2}]$ | $W_iW_j - W_jW_i$ <br> $W_iR_j - W_jW_i$ <br> $W_iW_j - R_jW_i$ <br> $W_iR_jC_i - W_jW_iC_i$ <br> $W_iR_jC_j - R_jC_jW_i$ |
| (8) | Lost Update | $R_i[x_m] \dots W_j[x_{m+1}] \dots W_i[x_{m+2}]$ | $R_iW_j - W_jW_i$ <br> $R_iW_j - R_jW_i$ <br> $R_iW_jC_i - W_jW_iC_i$ <br> $R_iW_jC_i - R_jW_iC_i$ <br> $R_iW_jC_j - R_jW_iC_j$ |
| (9) | Lost Self Update | $W_i[x_m] \dots W_j[x_{m+1}] \dots R_i[x_{m+1}]$ | $W_iW_j - W_jR_i$ <br> $W_iR_j - W_jR_i$ <br> $W_iW_jC_j - W_jR_iC_j$ <br> $W_iR_jC_j - W_jR_iC_j$ |
| (10) | Non-repeatable Read | $R_i[x_m] \dots W_j[x_{m+1}] \dots R_i[x_{m+1}]$ | $R_iW_j - W_jR_i$ <br> $R_iW_jC_j - W_jR_iC_j$ <br> $R_iW_jC_i - W_jR_iC_i$ |
| (11) | Intermediate Read | $W_i[x_m] \dots R_j[x_m] \dots W_i[x_{m+1}]$ | $W_iR_j - R_jW_i$ <br> $W_iR_jC_i - R_jW_iC_i$ |

By enumerating all single-variable 2-transaction cycles, we obtained 11 combinations of double-edge PG. In general, cases (1)(3)(4)(7)(8)(9) have $W_{i/j}[x_m] \dots W_{j/i}[x_{m+1}]$, while cases (2)(10)(11) don't. Instead, cases (2)(10)(11) have first-write then-read operation $W_i[x_m] \dots R_j[x_m]$ or $W_j[x_{m+1}] \dots R_i[x_{m+1}]$. Interestingly in (5) and (6), they use $C_j$ to divide $W_j[x_{m+1}] \dots R_i[x_{m+1}]$ and $W_{i/j}[x_m] \dots W_{j/i}[x_{m+1}]$, respectively.

Base on our classification of single-variable data anomalies, we then can classify data anomalies with an arbitrary number of transactions and variables. For a $PCG = (V, E)$, the set of variables in PCG is $D_{PCG}(s)$, then we define:

**1) Write Anomaly Type (WAT):** If the PCG includes $W_{i/j}[x_m] \dots W_{j/i}[x_{m+1}]$, i.e.,
$$E_{WAT} = \{PCG(V,E)| \text{ exists } W_i[x_m] \dots W_j[x_{m+1}] \text{ or } W_j[x_m] \dots W_i[x_{m+1}]\}$$
Particularly, when $D(s) = \{x\}$, data anomalies (1)(3)(4)(7)(8)(9)$\in E_{WAT}[x]$. In this case, we call it WAT-SDA, where SDA is short for Single-variable Double-transaction Anomaly.

**2) Read Anomaly Type (RAT):** If the PCG can includes $W_{i/j}[x_m] \dots R_{j/i}[x_m]$ when PCG does not include $W_{i/j}[x_m] \dots W_{j/i}[x_{m+1}]$, i.e.,
$$E_{RAT} = \{PCG(V,E)| \text{not exists } W_{i/j}[x_m] \dots W_{j/i}[x_{m+1}] \wedge \text{ exists } W_{i/j}[x_m] \dots R_{j/i}[x_m]\}$$
Particularly, when $D(s) = \{x\}$, data anomalies (2)(10)(11)$\in E_{RAT}[x]$. In this case, we call it RAT-SDA.

**3) Intersect Anomaly Type (IAT):** PCG that are not the above cases, i.e.,
$$E_{IAT} = \{PCG(V,E)| \text{ not exists } W_{i/j}[x_m] \dots W_{j/i}[x_{m+1}] \wedge \text{ not exists} W_{i/j}[x_m] \dots R_{j/i}[x_m]\}$$
Particularly, when $D_{PCG}(s) = \{x\}$, data anomalies (5)(6)$\in E_{RAT}[x]$. In this case, we call it IAT-SDA.

### 3.3.2 Classification of Double-variable Double-transaction Data Anomalies

Generally, for any number of variables, we can construct the PCG and simplify it to formal expression as well as obtain the category of data anomalies. In this part, we will divide the bivariable 2-transaction data anomalies.

We still enumerate all the cases of double-variable double-transaction cycles. Let $CG(s) = (\{t_i, t_j\}, \{p_{ij}[x], p_{ji}[y]\})$ be the DDA. By $t_i <_s t_j$, there are 15 forms of data anomalies that can be constructed (see Appendix 2 for anomalies (12) - (26)), as summarizes in Table

4.

Table 4 PCGs with double variables and two edges.

| No | Anomaly name | Formal expression | PCG |
|---|---|---|---|
| (12) | Double-Write Skew2 Committed | $W_i[x_m] \ldots W_j[x_{m+1}] \ldots W_j[y_n] \ldots C_j \ldots R_i[y_n]$ | $W_iW_jC_j[x] - W_jC_jR_i[y]$ |
| (13) | Full-Write Skew Committed | $W_i[x_m] \ldots W_j[x_{m+1}] \ldots W_j[y_n] \ldots C_j \ldots W_i[y_{n+1}]$ | $W_iW_jC_j[x] - W_jC_jW_i[y]$ |
| (14) | Write-Read Skew Committed | $W_i[x_m] \ldots R_j[x_m] \ldots W_j[y_n] \ldots C_j \ldots R_i[y_n]$ | $W_iR_jC_j[x] - W_jC_jR_i[y]$ |
| (15) | Double-Write Skew1 Committed | $W_i[x_m] \ldots R_j[x_m] \ldots W_j[y_n] \ldots C_j \ldots W_i[y_{n+1}]$ | $W_iR_jC_j[x] - W_jC_jW_i[y]$ |
| (16) | Read Skew Committed | $R_i[x_m] \ldots W_j[x_{m+1}] \ldots W_j[y_n] \ldots C_j \ldots R_i[y_n]$ | $R_iW_jC_j[x] - W_jC_jR_i[y]$ |
| (17) | Read-Write Skew1 Committed | $R_i[x_m] \ldots W_j[x_{m+1}] \ldots W_j[y_n] \ldots C_j \ldots W_i[y_{n+1}]$ | $R_iW_jC_j[x] - W_jC_jW_i[y]$ |
| (18) | Full Write Skew | $W_i[x_m] \ldots W_j[x_{m+1}] \ldots W_j[y_n] \ldots W_i[y_{n+1}]$ | $W_iW_j[x] - W_jW_i[y]$ |
| (19) | Double-Write Skew1 | $W_i[x_m] \ldots R_j[x_m] \ldots W_j[y_n] \ldots W_i[y_{n+1}]$ | $W_iR_j[x] - W_jW_i[y]$<br>$W_iR_jC_i[x] - W_jW_iC_i[y]$ |
| (20) | Read-Write Skew1 | $R_i[x_m] \ldots W_j[x_{m+1}] \ldots W_j[y_n] \ldots W_i[y_{n+1}]$ | $R_iW_j[x] - W_jW_i[y]$<br>$R_iW_jC_i[x] - W_jW_iC_i[y]$ |
| (21) | Double-Write Skew2 | $W_i[x_m] \ldots W_j[x_{m+1}] \ldots W_j[y_n] \ldots R_i[y_{n+1}]$ | $W_iW_j[x] - W_jR_i[y]$<br>$W_iW_jC_j[x] - W_jR_iC_j[y]$ |
| (22) | Write-Read Skew | $W_i[x_m] \ldots R_j[x_m] \ldots W_j[y_n] \ldots R_i[y_n]$ | $W_iR_j[x] - W_jR_i[y]$<br>$W_iR_jC_i[x] - W_jR_iC_i[y]$<br>$W_iR_jC_j[x] - W_jR_iC_j[y]$ |
| (23) | Read Skew | $R_i[x_m] \ldots W_j[x_{m+1}] \ldots W_j[y_n] \ldots R_i[y_n]$ | $R_iW_j[x] - W_jR_i[y]$<br>$R_iW_jC_j[x] - W_jR_iC_j[y]$<br>$R_iW_jC_i[x] - W_jR_iC_i[x]$ |
| (24) | Read-Write Skew2 | $W_i[x_m] \ldots W_j[x_{m+1}] \ldots R_j[y_n] \ldots W_i[y_{n+1}]$ | $W_iW_j[x] - R_jW_i[y]$<br>$W_iW_jC_j[x] - R_jC_jW_i[y]$<br>$W_iW_jC_j[x] - R_jW_iC_j[y]$<br>$W_iW_jA_j[x] - R_jW_iA_j[y]$ |
| (25) | Read Skew2 | $W_i[x_m] \ldots R_j[x_m] \ldots R_j[y_n] \ldots W_i[y_{n+1}]$ | $W_iR_j[x] - R_jW_i[y]$<br>$W_iR_j[x] - R_jC_jW_i[y]$<br>$W_iR_jC_j[x] - R_jW_iC_j[y]$ |
| (26) | Write Skew | $R_i[x_m] \ldots W_j[x_{m+1}] \ldots R_j[y_n] \ldots W_i[y_{n+1}]$ | $R_iW_j[x] - R_jW_i[y]$<br>$R_iW_jC_j[x] - R_jC_jW_i[y]$<br>$R_iW_jC_j[x] - R_jW_iC_j[y]$<br>$R_iW_jA_j[x] - R_jW_iA_j[y]$ |

We now can classify the double-variable double-transaction data anomalies. Firstly, (12)(13)(18)-(21)(24)(25) $\in E_{WAT}$, we call it WAT-DDA, where DDA is short for Double-variables by Double-transactions Anomalies. And, (14)(15)(22)(23) $\in E_{RAT}$, we call it RAT-DDA. The rest cases (16)(17)(26) $\in E_{IAT}$, named IAT-DDA.

Example 7: We show how we redefine the know data anomalies Lost Update and Read Skew in Table 5:

Table 5    Examples of New Definition of Known Data Anomalies

| Anomaly name | Formal expression | PCG |
|---|---|---|
| Lost Update Committed | $R_1[x_m] \ldots W_2[x_{m+1}] \ldots C_2 \ldots W_1[x_{m+2}]$ | Figure 10-(k)<br>$V = \{t_1, t_2\}$<br>$E = \{R_1W_2C_2[x], W_2C_2W_1[x]\}$ |
| Read Skew Committed | $R_1[x_m] \ldots W_2[x_{m+1}] \ldots W_2[y_n] \ldots C_2 \ldots R_1[y_n]$ | Figure 11-(m)<br>$V = \{t_1, t_2\}$<br>$E = \{R_1W_2C_2[x], W_2C_2R_1[y]\}$ |

以下我们讨论多事务多变量的偏序环,以下简称多事务异常.
In the following, we will discuss PCG with multiple variables and multiple transactions, i.e., $C_{da} = (N_D, N_T)$, where $N_D = |D(s)|, N_D \geq 2$ is the number of variables and $N_T = |T(s)|, N_T \geq 3$ is the number of transactions.

Similarly, for PCG with multiple variables and multiple transactions, if they meet:

1) $W_{i/j}[x_m]…W_{j/i}[x_{m+1}]$ exists in PCG, then we call it WAT-MDA, where MDA is short for Multi-variable Double-transaction Anomalies；
2) $W_{i/j}[x_m]…W_{j/i}[x_{m+1}]$ does not exist in PCG, but $W_{i/j}[x_m]…R_{j/i}[x_m]$ exists in PCG, we call it RAT-MDA;
3) PCG that are not the above cases, we call it IAT-MDA.

We summarize the data anomalies in Figure 9.

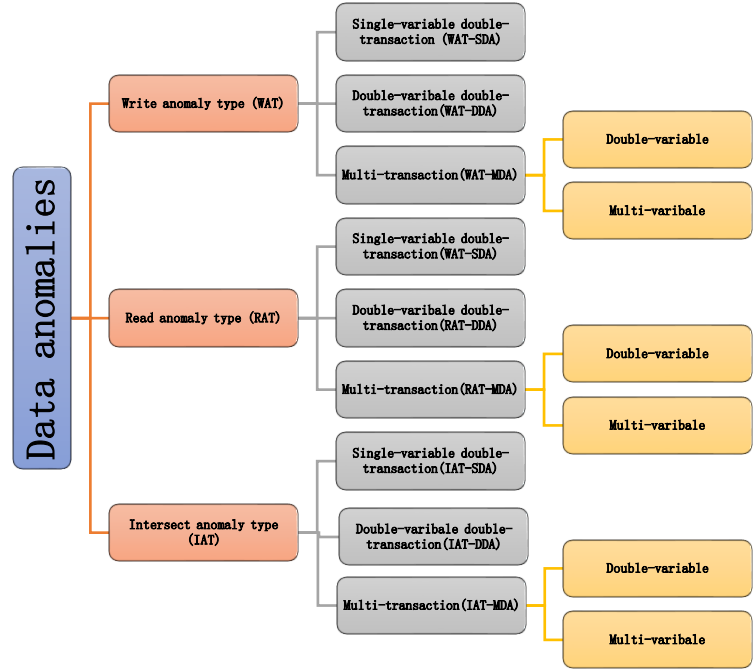

**Fig. 9**   Classification summary of data anomalies.

Based on the above definition and classification, we summarize every data anomaly in Table 6. And these data anomalies, we rename and formalize them as well as provide graphic presentation accordingly.

**Table 6**   Definition and Classification of Data Anomalies (Bold font, newly reported data anomalies)

| Class | Sub-class/$C_{da}$ | Anomaly name | Formal expression | No | PCG representation |
|---|---|---|---|---|---|
| WAT | SDA | Dirty Write | $W_i[x_m]…W_j[x_{m+1}]…A_i/C_i$ | (1) | Figure 10-(a) |
| | SDA | **Lost Self Update Committed** | $W_i[x_m]…W_j[x_{m+1}]…C_j…R_i[x_{m+1}]$ | **(3)** | **Figure 10-(b)** |
| | SDA | **Full-Write Committed** | $W_i[x_m]…W_j[x_{m+1}]…C_j…W_i[x_{m+2}]$ | **(4)** | **Figure 10-(c)** |
| | SDA | **Full-Write** | $W_i[x_m]…W_j[x_{m+1}]…W_i[x_{m+2}]$ | **(7)** | **Figure 10-(d)** |
| | SDA | Lost Update | $R_i[x_m]…W_j[x_{m+1}]…W_i[x_{m+2}]$ | (8) | Figure 10-(e) |
| | SDA | **Lost Self Update** | $W_i[x_m]…W_j[x_{m+1}]…R_i[x_{m+1}]$ | **(9)** | **Figure 10-(f)** |
| | DDA | **Double-Write Skew 2 Committed** | $W_i[x_m]…W_j[x_{m+1}]…W_j[y_n]…C_j…R_i[y_n]$ | **(12)** | **Figure 11-(a)** |
| | DDA | **Full-Write Skew Committed** | $W_i[x_m]…W_j[x_{m+1}]…W_j[y_n]…C_j…W_i[y_{n+1}]$ | **(13)** | **Figure 11-(b)** |
| | DDA | **Full-Write Skew** | $W_i[x_m]…W_j[x_{m+1}]…W_j[y_n]…W_i[y_{n+1}]$ | **(18)** | **Figure 11-(c)** |
| | DDA | **Double-Write Skew 1** | $W_i[x_m]…R_j[x_m]…W_j[y_n]…W_i[y_{n+1}]$ | **(19)** | **Figure 11-(d)** |
| | DDA | **Double-Write Skew 2** | $W_i[x_m]…R_j[x_m]…W_j[y_n]…W_i[y_{n+1}]$ | **(21)** | **Figure 11-(f)** |
| | DDA | **Read-Write Skew 1** | $R_i[x_m]…W_j[x_{m+1}]…W_j[y_n]…W_i[y_{n+1}]$ | **(20)** | **Figure 11-(e)** |
| | DDA | **Read-Write Skew 2** | $W_i[x_m]…W_j[x_{m+1}]…R_j[y_n]…W_i[y_{n+1}]$ | **(24)** | **Figure 11-(g)** |
| | MDA | **Step WAT** | $…W_{i/j}[x_m]…W_{j/i}[x_{m+1}]…$, and $N_D \geqslant 3$ | | **Figure 12-(a)** |
| RAT | SDA | Dirty Read | $W_i[x_m]…R_j[x_m]…A_i$ | (2) | Figure 10-(g) |
| | SDA | Non-repeatable Read | $R_i[x_m]…W_j[x_{m+1}]…R_i[x_{m+1}]$ | (10) | Figure 10-(h) |
| | SDA | Intermediate Read | $W_i[x_m]…R_j[x_m]…W_i[x_{m+1}]$ | (11) | Figure 10-(i) |
| | DDA | Write-Read Skew Committed | $W_i[x_m]...R_j[x_m]...W_j[y_n]...C_j...R_i[y_n]$ | (14) | Figure 11-(h) |
| | DDA | Double-Write Skew 1 Committed | $W_i[x_m]...R_j[x_m]...W_j[y_n]...C_j...W_i[y_{n+1}]$ | (15) | Figure 11-(i) |
| | DDA | Write-Read Skew | $W_i[x_m]...R_i[x_m]...W_j[y_n]...R_i[y_n]$ | (22) | Figure 11-(j) |
| | DDA | Read Skew | $R_i[x_m]...W_j[x_{m+1}]...W_j[y_n]...R_i[y_n]$ | (23) | Figure 11-(k) |
| | DDA | Read Skew 2 | $W_i[x_m]...R_j[x_m]...R_j[y_n]...W_i[y_{n+1}]$ | (25) | Figure 11-(l) |
| | MDA | Step RAT | $...W_{i/j}[x_m]...R_{j/i}[x_m]...$, and $N_D \geq 3$ and not include $(...W_{i/j}[x_m]...W_{j/i}[x_{m+1}]...)$ | | Figure 12-(b) |
| IAT | SDA | Non-repeatable Read Committed | $R_i[x_m]...W_j[x_{m+1}]...C_j...R_i[x_{m+1}]$ | (5) | Figure 10-(j) |
| | SDA | Lost Update Committed | $R_i[x_m]...W_j[x_{m+1}]...C_j...W_i[x_{m+2}]$ | (6) | Figure 10-(k) |
| | DDA | Read Skew Committed | $R_i[x_m]...W_j[x_{m+1}]...W_j[y_n]...C_j...R_i[y_n]$ | (16) | Figure11-(m) |
| | DDA | Read-Write Skew 1 Committed | $R_i[x_m]...W_j[x_{m+1}]...W_j[y_n]...C_j...W_i[y_{n+1}]$ | (17) | Figure 11-(n) |
| | DDA | Write Skew | $R_i[x_m]...W_j[x_{m+1}]...R_j[y_n]...W_i[y_{n+1}]$ | (26) | Figure 11-(o) |
| | MDA | Step IAT | Not include $(...W_{i/j}[x_m]...R_{j/i}[x_m]...$ and $...W_{i/j}[x_m]...W_{j/i}[x_{m+1}]...)$ $N_D \geq 3$ | | Figure 12-(c) |

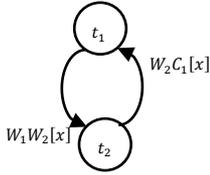

$s_1 = R_1[x_0]W_1[x_0]W_2[x_1]A_1$
$Pop(s_1) = \{(W_1W_2A_1[x])\}$

(a)  Dirty Write

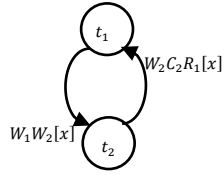

$s_2 = W_1[x_1]W_2[x_2]C_2R_1[x_2]$
$Pop(s_2) = \{(W_1W_2C_2[x]), (W_2C_2R_1[x])\}$

(b)  Lost Self Update Committed

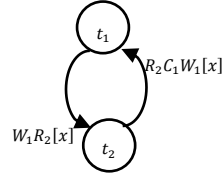

$s_3 = W_1[x_1]R_2[x_1]C_2W_1[x_2]$
$Pop(s_3) = \{(W_1R_2C_2[x]), (R_2C_2W_1[x])\}$

(c)  Full-Write Committed

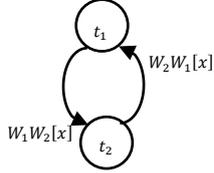

$s_4 = W_1[x_1]W_2[x_2]W_1[x_3]$
$Pop(s_4) = \{(W_1W_2[x]), (W_2W_1[x])\}$

(d)  Full-Write

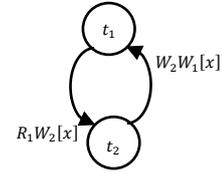

$s_5 = R_1[x_0]W_2[x_1]W_1[x_2]$
$Pop(s_5) = \{(R_1W_2[x]), (W_2W_1[x])\}$

(e)  Lost Update

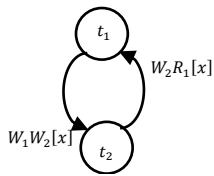

$s_6 = W_1[x_1]W_2[x_2]R_1[x_2]$
$Pop(s_6) = \{(W_1W_2[x]), (W_2R_1[x])\}$

(f)  Lost Self Update

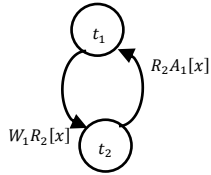

$s_7 = W_1[x_1]R_2[x_1]A_1$
$Pop(s_7) = \{(W_1R_2A_1[x])\}$

(g)  Dirty Read

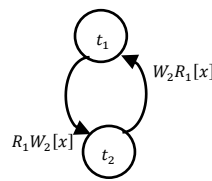

$s_8 = R_1[x_0]W_2[x_1]R_1[x_1]$
$Pop(s_8) = \{(R_1W_2[x]), (W_2R_1[x])\}$

(h)  Non-repeatable Read

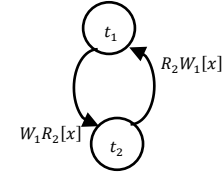

$s_9 = W_1[x_1]R_2[x_1]W_1[x_2]$
$Pop(s_9) = \{(W_1R_2[x]), (R_2W_1[x])\}$

(i)  Intermediate Read

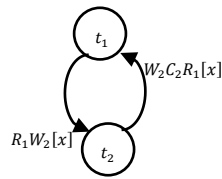

$s_{10} = R_1[x_0]W_2[x_1]C_2R_1[x_1]$
$Pop(s_{10}) = \{(R_1W_2C_2[x]), (W_2C_2R_1[x])\}$

(j)  Non-repeatable Read Committed

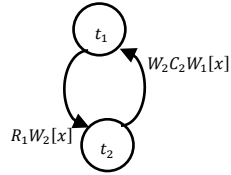

$s_{11} = R_1[x_0]W_2[x_1]C_2W_1[x_2]$
$Pop(s_{11}) = \{(R_1W_2C_2[x]), (W_2C_2W_1[x])\}$

(k)  Lost Update Committed

**Fig.10** PCG of SDA

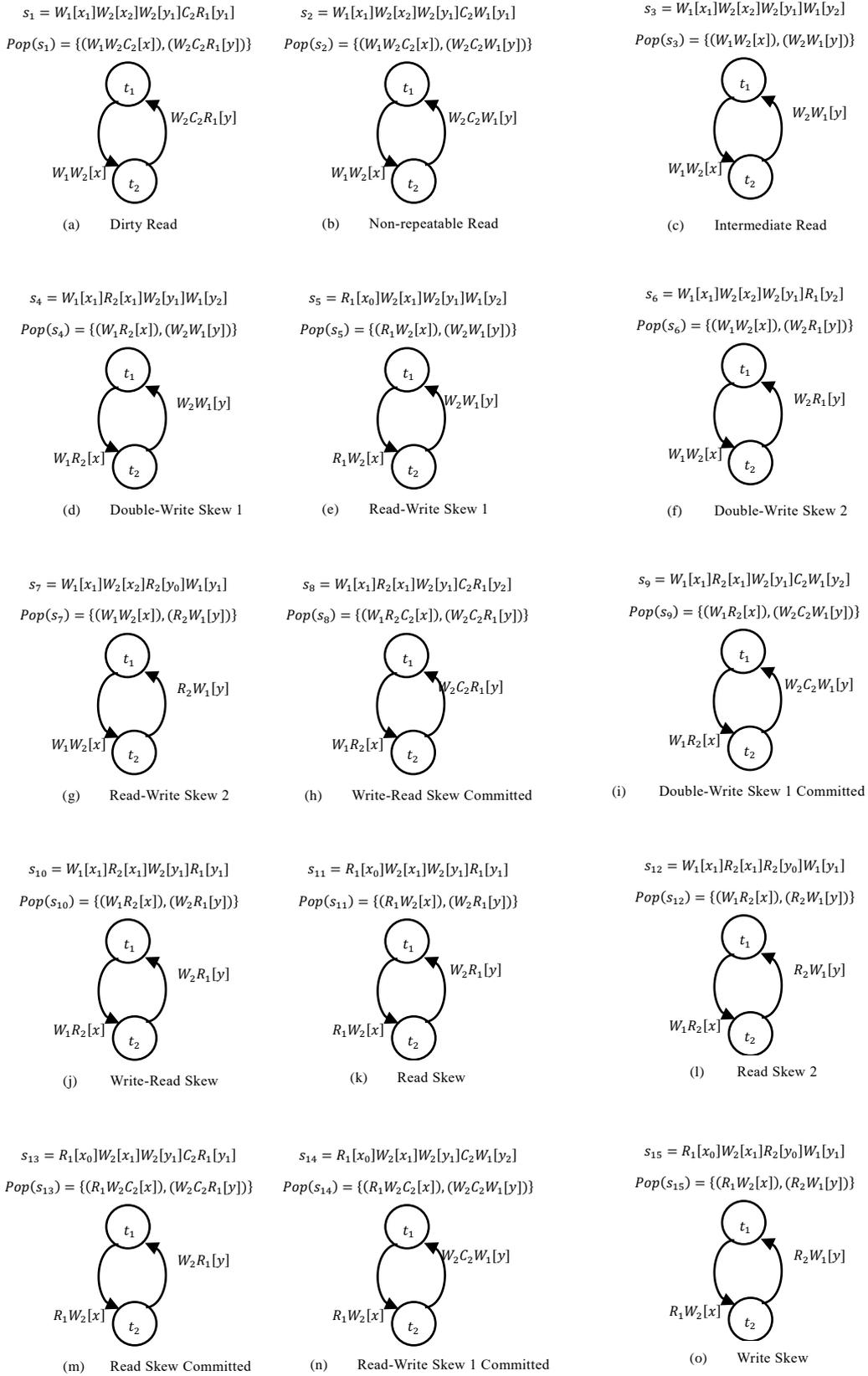

**Fig.11** PCG of DDA

Example 8: We show some possible schedules for MDA in Figure 9:

1. Figure 12-a depicts Step WAT with schedule $s = [x_0]W_2[x_1]W_2[y_1]W_3[y_2]R_3[z_0]W_1[z_1]$. $WW$ exists in schedule $s$.
2. Figure 12-b depicts Step RAT with schedule $s = R_1[x_0]W_2[x_1]W_2[y_1]R_3[y_1]R_3[z_0]W_1[z_1]$. $WW$ does not exist in schedule $s$, and $WR$ exists in schedule $s$.
3. Figure 12-c depicts Step IAT with schedule $s = R_1[x_0]W_2[x_1]R_2[y_0]W_3[y_1]R_3[z_0]W_1[z_1]$. $WW$ and $WR$ do not exist in schedule $s$, and $RW$ exists in schedule $s$. Step IAT consists of $\{RW、WCR、WCW\}$, and it must include $RW$.

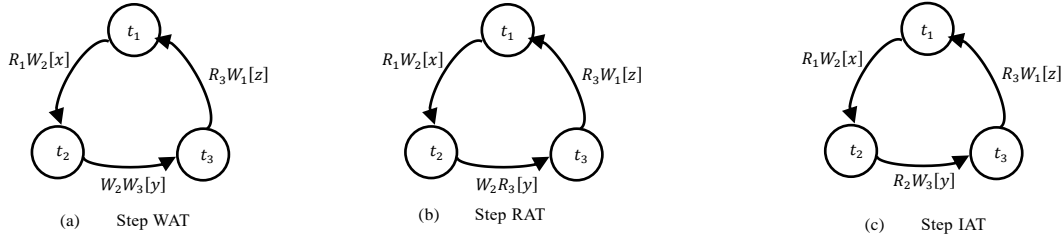

Fig.12   PCG of MDA

## 4 Isolation Levels

Most isolation levels [16,18,20] are defined based on a limited number of known data anomalies, making it difficult to understand the data anomalies. The reason firstly is lacking a systematical definition of data anomalies. The isolation levels defined by these know data anomalies are difficult to capture unknown (Table 1) and infinite data anomalies (we show in Section 3.2). Secondly, different isolation levels have their own standards. For example, some [16,18] define isolation based on the limited anomaly cases, while some, e.g., [20] define based on the characteristics of conflict cycle and specific data anomalies. Finally, isolation levels can be complex with many categories, e.g., four levels in [16] and eight levels in [18].

The above isolation levels use different levels to avoid different data anomalies. They essentially classify the data anomalies. The definition of isolation levels itself in fact is a better way to classify data anomalies. Traditional isolation levels divide several levels with a hierarchical structure such that higher level also avoids lower level anomalies. This is only one way of classification. So, based on different situations, we may flexibly define different but more suitable levels, to better practical deployment or better design of concurrent control algorithms.

Based on the definition and classification of data anomalies in Table 6, we provide two means of isolation level definitions. In both cases, we cover all data anomalies, and they are flexible yet integrated levels

Firstly, we define simplified isolation levels, which only have two levels shown in Table 7. A simple definition of isolation levels can avoid a large number of data anomalies at a low cost. It also may simplify the design of concurrent control algorithms. It is easy to be applied in production:

1. No Read-Write Anomalies(NRW): In this level, it does not allow WAT and RAT anomalies, but may allow anomalies.
2. No Anomalies(NA): In this level, no anomalies are allowed, which is equivalent to the serializable isolation level in the ANSI-SQL standard[16].
3. NRW is a weaker level, NA is the strictest level.

As discussed, a simplified isolation levels may help better design of concurrent control algorithm. For example, based on Table 7, we may use rules to achieve NRW level, e.g., by avoiding WW in WAT anomalies or avoiding RAT via read committed with snapshot techniques. And to obtain NA level, we may only need to identify two continuous RW in serializable snapshot isolation. These simplified isolation levels may make concurrent control algorithms become concise but efficient. Using rules to avoid many anomalies is a low-cost implementation and may achieve high throughput with efficient and concise validation. These strategy may also apply to other concurrent control algorithm like 2PL[1],T/O[12],OCC[27],MVCC[14,28,29].

Table 7   Simplified isolation levels for engineering practice

| Class | Sub class/$C_{da}$ | Anomaly name | NRW | NA |
|---|---|---|---|---|
| WAT | SDA | Dirty Write | Not Possible | Not Possible |
| | SDA | Lost Update | Not Possible | Not Possible |
| | SDA | Lost Self Update | Not Possible | Not Possible |
| | SDA | Full-Write | Not Possible | Not Possible |
| | SDA | Full-Write Committed | Not Possible | Not Possible |
| | DDA | Read-Write Skew 1 | Not Possible | Not Possible |

| Class | Sub Class/$C_{da}$ | Anomaly name | | |
|---|---|---|---|---|
| | DDA | Read-Write Skew 2 | Not Possible | Not Possible |
| | DDA | Double-Write Skew 1 | Not Possible | Not Possible |
| | DDA | Double-Write Skew 2 | Not Possible | Not Possible |
| | DDA | Double-Write Skew 2 Committed | Not Possible | Not Possible |
| | DDA | Full-Write Skew | Not Possible | Not Possible |
| | DDA | Full-Write Skew Committed | Not Possible | Not Possible |
| | MDA | Step WAT | Not Possible | Not Possible |
| RAT | SDA | Dirty Read | Not Possible | Not Possible |
| | SDA | Non-repeatable Read | Not Possible | Not Possible |
| | SDA | Non-repeatable Read Committed | Not Possible | Not Possible |
| | SDA | Intermediate Read | Not Possible | Not Possible |
| | DDA | Read Skew | Not Possible | Not Possible |
| | DDA | Read Skew 2 | Not Possible | Not Possible |
| | DDA | Write-Read Skew | Not Possible | Not Possible |
| | DDA | Write-Read Skew Committed | Not Possible | Not Possible |
| | DDA | Double-Write Skew 1 Committed | Not Possible | Not Possible |
| | MDA | Step RAT | Not Possible | Not Possible |
| IAT | SDA | Lost Update Committed | Possible | Not Possible |
| | DDA | Read Skew Committed | Possible | Not Possible |
| | DDA | Read-Write Skew 1 Committed | Possible | Not Possible |
| | DDA | Write Skew | Possible | Not Possible |
| | MDA | Step IAT | Possible | Not Possible |

Next, we provide a fine-grained definition of isolation levels, which consist of four levels shown in Table 8. The classification basis is that data anomalies can be avoided by a variety of simple rules. However, different rules are applicable to different levels, so that multiple isolation levels are needed. This method can be used in the teaching demonstration to show the subtle relation between data anomalies and isolation levels. We show the four isolation levels as follows:

1. No Write Anomalies(NW): In this level, it does not allow WAT anomalies, but may allow RAT and IAT anomalies.
2. No Read-Write Anomalies(NRW): In this level, it does not allow WAT and RAT, but may allow IAT, phantom read, and read skew anomalies.
3. No Predicate Anomalies(NPA): In this level, it does not allow WAT, RAT, phantom read, read skew, but may allow IAT. Worth mentioning in this paper, we have not discussed the anomalies with predicates for simplicity. So there are no definitions and descriptions for phantom read in Tables 6, 7, and 8.
4. No Anomalies(NA): In this level, no anomalies are allowed, which is equivalent to the serializable isolation level in the ANSI-SQL standard[16].
5. NW is the weakest level, then the NRW. NPA is stronger than NRW, and NA is the strictest one.
6. For NW level, we can avoid anomalies by dealing with WW conflict. For NRW level, we can avoid anomalies by read-committed strategy. And for NPA, it can apply snapshot techniques. However, these strategies are not the only way to avoid anomalies. Each strategy may be used to satisfy an isolation level. Separating using the above three strategies is a way of instruction demonstration for isolation levels. And using these three strategies simultaneously is equivalent to NRW in a simplified isolation level. Therefore, isolation levels can be defined flexibly for different scenarios.

Table 8  Fine grained isolation level for Instruction Presentation

| Class | Sub Class/$C_{da}$ | Anomaly name | NW | NRW | NPA | NA |
|---|---|---|---|---|---|---|
| WAT | SDA | Dirty Write | Not Possible | Not Possible | Not Possible | Not Possible |
| | SDA | Lost Update | Not Possible | Not Possible | Not Possible | Not Possible |
| | **SDA** | **Lost Self Update** | Not Possible | Not Possible | Not Possible | Not Possible |
| | SDA | **Full-Write** | Not Possible | Not Possible | Not Possible | Not Possible |
| | SDA | **Full-Write Committed** | Not Possible | Not Possible | Not Possible | Not Possible |
| | **DDA** | **Read-Write Skew 1** | Not Possible | Not Possible | Not Possible | Not Possible |
| | DDA | **Read-Write Skew 2** | Not Possible | Not Possible | Not Possible | Not Possible |
| | DDA | **Double-Write Skew 1** | Not Possible | Not Possible | Not Possible | Not Possible |
| | DDA | **Double-Write Skew 2** | Not Possible | Not Possible | Not Possible | Not Possible |
| | DDA | **Double-Write Skew 2 Committed** | Not Possible | Not Possible | Not Possible | Not Possible |
| | DDA | **Full-Write Skew** | Not Possible | Not Possible | Not Possible | Not Possible |
| | DDA | **Full-Write Skew Committed** | Not Possible | Not Possible | Not Possible | Not Possible |
| | MDA | **Step WAT** | Not Possible | Not Possible | Not Possible | Not Possible |
| | SDA | Dirty Read | Possible | Not Possible | Not Possible | Not Possible |
| | SDA | Non-repeatable Read | Possible | Not Possible | Not Possible | Not Possible |
| | **SDA** | **Non-repeatable Read Committed** | Possible | Possible | Not Possible | Not Possible |
| | SDA | Intermediate Read | Possible | Not Possible | Not Possible | Not Possible |
| RAT | DDA | Read Skew | Possible | Possible | Not Possible | Not Possible |
| | DDA | **Read Skew 2** | Possible | Not Possible | Not Possible | Not Possible |
| | DDA | **Write-Read Skew** | Possible | Not Possible | Not Possible | Not Possible |
| | DDA | **Write-Read Skew Committed** | Possible | Not Possible | Not Possible | Not Possible |
| | DDA | **Double-Write Skew 1 Committed** | Possible | Not Possible | Not Possible | Not Possible |
| | MDA | **Step RAT** | Possible | Not Possible | Not Possible | Not Possible |
| | SDA | **Lost Update Committed** | Possible | Possible | Possible | Not Possible |
| | DDA | **Read Skew Committed** | Possible | Possible | Possible | Not Possible |
| IAT | DDA | **Read-Write Skew 1 Committed** | Possible | Possible | Possible | Not Possible |
| | DDA | Write Skew | Possible | Possible | Possible | Not Possible |
| | MDA | **Step IAT** | Possible | Possible | Possible | Not Possible |

## 5  Conclusion

This paper systematically studies data anomalies, and defines data anomalies by PCG. We classify PCG to classify all data anomalies. Then, concise isolation levels are defined based on the classification of data anomalies. For our future work, we would like to study anomalies with predicates, and see how these classification and isolation levels can help to design new concurrent control algorithms. Hopefully, we can contribute more to integrating the data anomalies research.

We would specially like to thank the cooperation and support from the Key Laboratory of Data Engineering and Knowledge Engineering of the Ministry of Education, Renmin University of China. We appreciate all the hard work from the authors and valuable comments from reviews.

In addition, we also implement some practical concurrent control algorithms, which are open source and available in Github (https://github.com/tencent/3TS).

**References**:

[1]  Kapali P. Eswaran,Jim Gray,Raymond A. Lorie,Irving L. traiger: the Notions of Consistency and Predicate Locks in a Database System. Commun. ACM 19(11): 624-633 (1976)

[2]  Jim Gray,Raymond A. Lorie,Gianfranco R. Putzolu,Irving L. traiger:Granularity of Locks and Degrees of Consistency in a Shared Data Base. IFIP Working Conference on Modelling in Data Base Management Systems 1976: 365-394

[3]  Andrea Cerone,Alexey Gotsman,and Hongseok Yang. Algebraic laws for weak consistency. International Conference on Concurrency theory (CONCUR 2017),LIPICS 85,pages 26:1-26:18,2017.

[4]  Andrea Cerone and Alexey Gotsman. Analysing snapshot isolation. Journal of the ACM (JACM 2018),65(2),11:1-11:41,2018.

[5]  Christos H. Papadimitriou: the serializability of concurrent database updates. J. ACM 26(4): 631-653 (1979)

[6]  P. A. Bernstein,D. Shipman,and W. Wong. Formal aspects of serializability in database concurrency control. IEEE transactions on Software Engineering,5(3):203–216,1979.

[7]  Alan Fekete,Shirley N. Goldrei,and Jorge Pérez Asenjo. Quantifying isolation anomalies. Proc. VLDB Endow.,2(1):467–478,August 2009.

[8]  Kamal Zellag and Bettina Kemme. Real-time quantification and classification of consistency anomalies in multi-tier architectures.

In Proceedings of the 27th IEEE International Conference on Data Engineering,ICDE '11,pages 613–624. IEEE Computer Society,2011

[9] Kamal Zellag and Bettina Kemme. How consistent is your cloud application? In Proceedings of the 3rd ACM Symposium on Cloud Computing,SoCC '12,pages 6:1–6:14. ACM,2012.

[10] Kamal Zellag and Bettina Kemme. Consistency anomalies in multi-tier architectures: Automatic detection and prevention. the VLDB Journal,3(1):147–172,2014

[11] Lamport L,Shostak R,Pease M. the Byzantine generals problem[J]. ACM transactions on Programming Languages and Systems (TOPLAS),1982,4(3): 382-401

[12] Roshan K. thomas,Ravi S. Sandhu: towards a Unified Framework and theory for Reasoning about Security and Correctness of transactions in Multilevel databases. DBSec 1993: 309-328.

[13] Kamal Zellag,Bettina Kemme: ConsAD: a real-time consistency anomalies detector. SIGMOD Conference 2012: 641-644

[14] Michael J. Cahill,Uwe Röhm,Alan David Fekete: Serializable isolation for snapshot databases. ACM trans. Database Syst. 34(4): 20:1-20:42 (2009)

[15] P. Bernstein,V. Hadzilacos,and N. Goodman. Concurrency Control and Recovery in Database Systems. Addison–Wesley,1987.

[16] ANSI X3.135-1992,American National Standard for Information Systems – Database Language – SQL,Nov 1992.

[17] Michael J. Cahill,Uwe Röhm,Alan David Fekete: Serializable isolation for snapshot databases. ACM trans. Database Syst. 34(4): 20:1-20:42 (2009)

[18] Hal Berenson,Philip A. Bernstein,Jim Gray,Jim Melton,Elizabeth J. O'Neil,Patrick E. O'Neil: A Critique of ANSI SQL Isolation Levels. SIGMOD Conference 1995: 1-10

[19] C. Xie,C. Su,C. Littley,L. Alvisi,M. Kapritsos,and Y. Wang. High-performance acid via modular concurrency control. In Proceedings of the 25th Symposium on Operating Systems Principles,SOSP'15,pages 279{294,New York,NY,USA,2015. ACM

[20] A. Adya,B. Liskov,and P. O'Neil. Generalized isolation level definitions. In Proceedings of the 16th International Conference on Data Engineering,ICDE '00,pages 67–78,Washington,DC,USA,2000. IEEE Computer Society.

[21] Ralf Schenkel,Gerhard Weikum,Norbert Weißenberg,et al. "Federated transaction Management with Snapshot Isolation". In: transactions and Database Dynamics. Vol. 1773. Springer Berlin / Heidelberg,2000,pp. 1–25.

[22] Alan Fekete,Elizabeth O'Neil,and Patrick O'Neil. A read-only transaction anomaly under snapshot isolation. SIGMOD Rec.,33(3):12–14,2004.

[23] A. Fekete,D. Liarokapis,E. O'Neil,P. O'Neil,and D. Shasha,"Making snapshot isolation serializable," ACM tODS,vol. 30,no. 2,pp. 492–528,2005.

[24] Ramez Elmasri,Shamkant B. Navathe: the Fundamentals of Database Systems,5th Edition

[25] Carsten Binnig,Stefan Hildenbrand,Franz Färber,Donald Kossmann,Juchang Lee,Norman May: Distributed snapshot isolation: global transactions pay globally,local transactions pay locally. VLDB J. 23(6): 987-1011 (2014)

[26] https://wiki.postgresql.org/wiki/SSI#Read_Only_transactions

[27] H. t. Kung,John t. Robinson: On Optimistic Methods for Concurrency Control. VLDB 1979: 351

[28] Christos H. Papadimitriou,Paris C. Kanellakis: On Concurrency Control by Multiple Versions. ACM trans. Database Syst. 9(1): 89-99 (1984)

[29] P. Bernstein,V. Hadzilacos,and N. Goodman. Concurrency Control and Recovery in Database Systems. Addison–Wesley,1987.

[30] Gerhard Weikum,Gottfried Vossen (2001): transactional Information Systems,Chapter 19,Elsevier,ISBN 1-55860-508-8

[31] P. Bailis,A. Fekete,J. M. Hellerstein,A. Ghodsi,and I. Stoica,"Scalable atomic visibility with ramp transactions," in SIGMOD,2014,pp. 27–38.

[32] Wang Shan, Sa Shixuan. Introduction to Database System (Fifth Edition). Higher education press, 2014

[33] Du Xiaoyong et al. Big data management. Higher education press, 2017

[34] Li HX, Li XY, Liu C, DU XY, Lu W,Pan AQ. Systematic definition and classification of data anomalies in DBMS. Ruan Jian Xue Bao/Journal of Software, 2021 (in Chinese). http://www.jos.org.cn/1000-9825/0000.htm

## Appendix 1: The proof of $W_i \ldots W_j \ldots (A_i \text{ or } C_i) \text{ and } (A_j \text{ or } C_j)$ anomalies

First, we define the operation of transaction status as follows:

- Let $W_i[x_n]$ denote the write operation of $t_i$, its variable version changing from $x_{n-1}$ to $x_n$;
- Let $A_i$ denote the operation to change from the current version to the original version of $t_i$;

> Let $C_i$ denote the operation to commit the current version.

Therefore, there are 8 combinations of $W_i \ldots W_j \ldots (A_i \text{ or } C_i) \text{ and } (A_j \text{ or } C_j)$, as follows:

1. $W_i[x_1] \ldots W_j[x_2] \ldots C_i \ldots C_j$:
   After operation $W_i[x_1] \ldots W_j[x_2]$, the version value of $x$ now is $x_2$. $C_i$ commits current version value $x = x_2$. However, if $t_i$ wants to commit with value $x = x_1$, then the anomaly occurs.
2. $W_i[x_1] \ldots W_j[x_2] \ldots C_i \ldots A_j$: Likewise in the case 1, the anomaly occurs when $C_i$ takes place;
3. $W_i[x_1] \ldots W_j[x_2] \ldots A_i \ldots C_j$:
   Since after operation $A_i$ takes place, the version value is $x = x_0$. $C_j$ commits current version value $x = x_0$. However, $t_j$ wants to commit with $x = x_2$, then the anomaly occurs.
4. $W_i[x_1] \ldots W_j[x_2] \ldots A_i \ldots A_j$
   Since after operation $A_i$ takes place, the version value is $x = x_0$. And after operation $A_j$ takes place, the version value is $x = x_1$. However, $x_1$ is never a committed version of database, meaning anomaly occurs.
5. $W_i[x_1] \ldots W_j[x_2] \ldots C_j \ldots C_i$
   $C_j$ commits with $x = x_2$, $C_i$ commits with $x = x_2$, however, $t_i$ wants to commit with $x = x_1$, then anomaly occurs.
6. $W_i[x_1] \ldots W_j[x_2] \ldots C_j \ldots A_i$
   $C_j$ commit with $x = x_2$. $A_i$ restores $t_i$'s original value $x = x_0$, the anomaly does not occur.
7. $W_i[x_1] \ldots W_j[x_2] \ldots A_j \ldots C_i$
   $A_j$ restores $t_j$'s original value $x = x_1$. $C_i$ commits with $x = x_1$. No data anomaly occurs;
8. $W_i[x_1] \ldots W_j[x_2] \ldots A_j \ldots A_i$
   $A_j$ restores $t_j$'s original value $x = x_1$. $A_i$ restores $t_i$'s original value $x = x_0$. No data anomaly occurs.

To sum up, $W_i[x_1] \ldots W_j[x_2] \ldots C_i$(cases 1-2) and $W_i[x_1] \ldots W_j[x_2] \ldots A_i$(cases 3-4) exist data anomalies; $W_i[x_1] \ldots W_j[x_2] \ldots C_j$ may exist depending on the subsequent operations. Interestingly, $W_i[x_1] \ldots W_j[x_2] \ldots A_j$ do not yield any anomalies.

Appendix 2: The proof of the PCG with single variable and two edges.

In total, there are 9 combinations of POP, i.e., $Pop = \{W_iC_iR_j, W_iC_iW_j, R_iC_iW_j, W_iW_j, W_iR_j, R_iW_j, W_iR_jA_i, W_iW_jC_i, W_iW_jA_i\}$. We can divide them into three categories, i.e., $W_iW_j = \{W_iW_j, W_iW_jC_j\}$, $W_iR_j = \{W_iR_j, W_iR_jC_j, W_iR_jC_i\}$, $R_iW_j = \{R_iW_j, R_iW_jC_j, R_iW_jC_i, R_iW_jA_i\}$. We show the proof in the follow.

When the $Pop(s)$ consists of $\{W_iW_jA_i, W_iW_jC_i, R_iW_jA_i\}$ in the schedule $s$, they themselves are PCG.

> $WA$ strands for PCG $W_i[x_m]W_j[x_{m+1}] - W_j[x_{m+1}]A_i$
   Its formal expression is $W_i[x_m] \ldots W_j[x_{m+1}] \ldots A_i$, i.e., Dirty Write.  (1)
> $WC$ stands for PCG $W_i[x_m]W_j[x_{m+1}] - W_j[x_{m+1}]C_i$
   Its formal expression is $W_i[x_m] \ldots W_j[x_{m+1}] \ldots C_i$, i.e., Dirty Write, likewise in (1).
> $RA$ stands for PCG $W_i[x_m]R_j[x_m] - R_j[x_m]A_i$
   Its formal expression is $W_i[x_m] \ldots R_j[x_m] \ldots A_i$, i.e., Dirty Read.  (2)

We now consider the combinational situation of $\{W_iC_iR_j, W_iC_iW_j, R_iC_iW_j, W_iW_j, W_iR_j, R_iW_j\}$.

For single variable double-transaction $PCG(s) = (\{t_i, t_j\}, \{p_{ij}[x], p_{ji}[x]\})$, by $t_i <_s t_j$, the possible combinational PCG are as follows:

1. $p_{ij}$ can not be $W_iC_iR_j$, $W_iC_iW_j$, or $R_iC_iW_j$. Otherwise, by $t_i <_s t_j$, $p_{ji}$ does not have the operation of $t_i$;
2. If $p_{ji}[x]$ exists $W_jC_jR_i$, $W_jC_jW_i$, or $R_jC_jW_i$, then $p_{ij}$ must have $C_i$, then the combinations can be the following cases:
   > $W_iW_jC_j - W_jC_jR_i$ or $W_iR_jC_j - W_jC_jR_i$
      Its formal expression is $W_i[x_m] \ldots W_j[x_{m+1}] \ldots C_j \ldots R_i[x_{m+1}]$, i.e., Lost Self Update Committed.  (3)
   > $W_iW_jC_j - W_jC_jW_i$ or $W_iR_jC_j - W_jC_jW_i$ or $W_iW_jC_j - R_jC_jW_i$
      Its formal expression is $W_i[x_m] \ldots W_j[x_{m+1}] \ldots C_j \ldots W_i[x_{m+2}]$, i.e., Full-Write Committed.  (4)
   > $R_iW_jC_j - W_jC_jR_i$
      Its formal expression is $R_i[x_m] \ldots W_j[x_{m+1}] \ldots C_j \ldots R_i[x_{m+1}]$, i.e., Non-repeatable Read Committed.  (5)
   > $R_iW_jC_j - W_jC_jW_i$ or $R_iW_jC_j - R_jC_jW_i$
      Its formal expression is $R_i[x_m] \ldots W_j[x_{m+1}] \ldots C_j \ldots W_i[x_{m+2}]$, i.e., Lost Update Committed.  (6)
   > $W_iR_jC_j - R_jC_jW_i$, this time, it does not exist the anomaly.
3. Based on the definition of POP, $W_iW_j = \{W_iW_j, W_iW_jC_j\}$, and the possible PCG by this POP are in the following.

1） If $p_{ji}[x] = W_jW_i$, then it will be the PCG.
- $W_iW_j - W_jW_i$ or $W_iR_j - W_jW_i$
  Its formal expression is $W_i[x_m] \ldots W_j[x_{m+1}] \ldots W_i[x_{m+2}]$, i.e., Full Write. (7)
- $R_iW_j - W_jW_i$
  Its formal expression is $R_i[x_m] \ldots W_j[x_{m+1}] \ldots W_i[x_{m+2}]$, i.e., Lost Update. (8)

2） If $p_{ji}[x] = W_jW_iC_i$, then $p_{ij}[x]$ must exist $C_i$, therefor exists a PCG.
- $W_iW_jC_i - W_jW_iC_i$, i.e., $p_{ij}[x] = W_iW_jC_i$ is a PCG similar to (1).
- $W_iR_jC_i - W_jW_iC_i$
  Its formal expression is $W_i[x_m] \ldots W_j[x_{m+1}] \ldots W_i[x_{m+2}]$, similar to (7).
- $R_iW_jC_i - W_jW_iC_i$, i.e., $p_{ij}[x] = R_iW_jC_i$, similar to (8).

4. Based on the definition of POP, $WR = \{W_iR_j, W_iR_jC_j, W_iR_jC_i\}$, and the possible PCG by this POP are in the following.

  1） If $p_{ji}[x] = W_jR_i$, then it will be the PCG.
  - $W_iW_j - W_jR_i$ or $W_iR_j - W_jR_i$
    Its formal expression is $W_i[x_m] \ldots W_j[x_{m+1}] \ldots R_i[x_{m+1}]$, i.e., Lost Self Update. (9)
  - $R_iW_j - W_jR_i$
    Its formal expression is $R_i[x_m] \ldots W_j[x_{m+1}] \ldots R_i[x_{m+1}]$, i.e., Non-repeatable Read. (10)

  2） If $p_{ji}[x] = W_jR_iC_i$, then $p_{ij}[x]$ must exist $C_i$, therefor exists a PCG.
  - $W_iW_jC_i - W_jR_iC_i$, i.e., $p_{ij}[x] = W_iW_jC_i$ is already a PCG, similar to (1).
  - $W_iR_jC_i - W_jR_iC_i$
    Its formal expression is $W_i[x_m] \ldots W_j[x_{m+1}] \ldots R_i[x_{m+1}]$, similar to (9).
  - $R_iW_jC_i - W_jR_iC_i$, i.e., $p_{ij}[x] = R_iW_jC_i$, similar to (10).

  3） If $p_{ji}[x] = W_jR_iC_j$, Then $p_{ij}[x]$ must exists $C_j$, therefor exists a PCG.
  - $W_iW_jC_j - W_jR_iC_j$
    Its formal expression is $W_i[x_m] \ldots W_j[x_{m+1}] \ldots R_i[x_{m+1}]$, similar to (9).
  - $W_iR_jC_j - W_jR_iC_j$
    Its formal expression is $W_i[x_m] \ldots W_j[x_{m+1}] \ldots R_i[x_{m+1}]$, similar to (9).
  - $R_iW_jC_j - W_jR_iC_j$,
    Its formal expression is $R_i[x_m] \ldots W_j[x_{m+1}] \ldots R_i[x_{m+1}]$, similar to (10).

5. According to the definition of POP, the POP $RW = \{R_iW_j, R_iW_jC_j, R_iW_jA_i\}$, may create following combinations:

  1） If $p_{ji}[x] = R_jW_i$, it may construct a PCG:
  - $W_iW_j - R_jW_i$
    Its formal expression is $W_i[x_m] \ldots W_j[x_{m+1}] \ldots W_i[x_{m+1}]$, similar to (7).
  - $W_iR_j - R_jW_i$
    Its formal expression is $W_i[x_m] \ldots R_j[x_m] \ldots W_i[x_{m+1}]$, i.e., Intermediate Read. (11)
  - $R_iW_j - R_jW_i$
    Its formal expression is $R_i[x_m] \ldots W_j[x_{m+1}] \ldots W_i[x_{m+2}]$, similar to (8).

  2） If $p_{ji}[x] = R_jW_iC_i$, then $p_{ij}[x]$ must exists $C_i$, therefor exists a PCG
  - $W_iW_jC_i - R_jW_iC_i$, now, $p_{ij}[x] = W_iW_jC_i$ has already formed a PCG, similar to (1).
  - $W_iR_jC_i - R_jW_iC_i$
    Its formal expression is $W_i[x_m] \ldots R_j[x_m] \ldots W_i[x_{m+1}]$, similar to (11).
  - $R_iW_jC_i - R_jW_iC_i$, i.e., $p_{ij}[x] = R_iW_jC_i$, similar to (8).

  3） If $p_{ji}[x] = R_jW_iC_j$, then $p_{ij}[x]$ must exist $C_j$, therefor exists a PCG.
  - $W_iW_jC_j - R_jW_iC_j$, i.e., $p_{ij}[x] = W_jW_iC_j$, therefore exists a PCG, similar to (1).
  - $W_iR_jC_j - R_jW_iC_j$, no anomaly occurs.
  - $R_iW_jC_j - R_jW_iC_j$, i.e., $p_{ij}[x] = R_iW_jC_i$, similar to (8).

  4） If $p_{ji}[x] = R_jW_iA_j$, then $p_{ij}[x]$ must exist $A_j$, therefor exists a PCG
  - $W_iW_jA_j - R_jW_iA_j$, there exists $W_j \ldots W_i \ldots A_j$, similar to (2);
  - $W_iR_jA_j - R_jW_iA_j$, no anomaly occurs;
  - $R_iW_jA_j - R_jW_iA_j$, there exists $W_j \ldots W_i \ldots A_j$, similar to (2);

## Appendix 3: The proof of double-variables double-edges PCG

Assume double-transaction double-variable $PCG(s) = (\{t_i, t_j\}, \{p_{ij}[x], p_{ji}[y]\})$. By $t_i <_s t_j$, the possible PCG can be the following:

1. With two variables, the set of POP in schedule $s$ may contain $\{W_iW_jA_i, W_iW_jC_i, R_iW_jA_i\}$. These POPs themselves are single-variable PCGs. So it would be the double-variable PCGs.
2. Likewise, $p_{ij}$ does not have $W_iC_iR_j$ or $W_iC_iW_j$. Otherwise, by $t_i <_s t_j$, $p_{ji}$ is not the operation in $t_i$.
3. If $p_{ji}[y]$ exists $W_jC_jR_i$ or $W_jC_jW_i$, them $p_{ij}$ must exists $C_j$, so the possible PCG are as follows:
    - $W_iW_jC_j[x] - W_jC_jR_i[y]$

Its formal expression is $W_i[x_m] \ldots W_j[x_{m+1}] \ldots W_j[y_n] \ldots C_j \ldots R_i[y_n]$, i.e., Double-Write Skew2 Committed.(12)
    - $W_iW_jC_j[x] - W_jC_jW_i[y]$

Its formal expression is $W_i[x_m] \ldots W_j[x_{m+1}] \ldots W_j[y_n] \ldots C_j \ldots W_i[y_{n+1}]$, i.e., Full-Write Skew Committed.(13)
    - $W_iR_jC_j[x] - W_jC_jR_i[y]$

Its formal expression is $W_i[x_m] \ldots R_j[x_m] \ldots W_j[y_n] \ldots C_j \ldots R_i[y_n]$, i.e., Write-Read Skew Committed. (14)
    - $W_iR_jC_j[x] - W_jC_jW_i[y]$

Its formal expression is $W_i[x_m] \ldots R_j[x_m] \ldots W_j[y_n] \ldots C_j \ldots W_i[y_{n+1}]$, i.e., Double-Write Skew1 Committed.(15)
    - $R_iW_jC_j[x] - W_jC_jR_i[y]$

Its formal expression is $R_i[x_m] \ldots W_j[x_{m+1}] \ldots W_j[y_n] \ldots C_j \ldots R_i[y_n]$, i.e., Read Skew Committed. (16)
    - $R_iW_jC_j[x] - W_jC_jW_i[y]$

Its formal expression is $R_i[x_m] \ldots W_j[x_{m+1}] \ldots W_j[y_n] \ldots C_j \ldots W_i[y_{n+1}]$, i.e., Read-Write Skew1 Committed.(17)

4. Based on the definition of POP, if $WW = \{W_iW_j, W_iW_jC_j\}$, then the possible PCG can be as follows:

    1) If $p_{ji}[y] = W_jW_i$, it can be in a PCG by:
    - $W_iW_j[x] - W_jW_i[y]$

Its formal expression is $W_i[x_m] \ldots W_j[x_{m+1}] \ldots W_j[y_n] \ldots W_i[y_{n+1}]$, i.e., Full Write Skew. (18)
    - $W_iR_j[x] - W_jW_i[y]$

Its formal expression is $W_i[x_m] \ldots R_j[x_m] \ldots W_j[y_n] \ldots W_i[y_{n+1}]$, i.e., Double-Write Skew1. (19)
    - $R_iW_j[x] - W_jW_i[y]$

Its formal expression is $R_i[x_m] \ldots W_j[x_{m+1}] \ldots W_j[y_n] \ldots W_i[y_{n+1}]$, i.e., Read-Write Skew1. (20)

    2) If $p_{ji}[y] = W_jW_iC_i$, then $p_{ij}[x]$ must exists $C_i$, therefor exists a PCG
    - $W_iW_jC_i[x] - W_jW_iC_i[y]$, i.e., $p_{ij}[x] = W_iW_jC_i$, therefor exists a PCG, similar to (1).
    - $W_iR_jC_i[x] - W_jW_iC_i[y]$

Its formal expression is $W_i[x_m] \ldots R_j[x_m] \ldots W_j[y_n] \ldots W_i[y_{n+1}]$, similar to (19).
    - $R_iW_jC_i[x] - W_jW_iC_i[y]$,此时$p_{ij}[x] = R_iW_jC_i$, similar to (20).

5. Based on the definition of POP, if $WR = \{W_iR_j, W_iR_jC_j, W_iR_jC_i\}$, then the possible PCG can be as follows:

    1) If $p_{ji}[y] = W_jR_i$, it can be in a PCG by:
    - $W_iW_j[x] - W_jR_i[y]$

Its formal expression is $W_i[x_m] \ldots W_j[x_{m+1}] \ldots W_j[y_n] \ldots R_i[y_{n+1}]$, i.e., Double-Write Skew2. (21)
    - $W_iR_j[x] - W_jR_i[y]$

Its formal expression is $W_i[x_m] \ldots R_j[x_m] \ldots W_j[y_n] \ldots R_i[y_n]$, i.e., Write-Read Skew. (22)
    - $R_iW_j[x] - W_jR_i[y]$

Its formal expression is $R_i[x_m] \ldots W_j[x_{m+1}] \ldots W_j[y_n] \ldots R_i[y_n]$, i.e., Read Skew. (23)

    2) If $p_{ji}[y] = W_jR_iC_i$, then $p_{ij}[x]$ must exists $C_i$, therefor exists a PCG
    - $W_iW_jC_i[x] - W_jR_iC_i[y]$, i.e., $p_{ij}[x] = W_iW_jC_i$, therefor exists a PCG, similar to (1).
    - $W_iR_jC_i[x] - W_jR_iC_i[y]$

Its formal expression is $W_i[x_m] \ldots R_j[x_m] \ldots W_j[y_n] \ldots R_i[y_n]$, similar to (22).
    - $R_iW_jC_i[x] - W_jR_iC_i[x]$,此时$p_{ij}[x] = R_iW_jC_i$, similar to (23).

    3) If $p_{ji}[y] = W_jR_iC_j$, then $p_{ij}[x]$ must exists $C_j$, therefor exists a PCG
    - $W_iW_jC_j[x] - W_jR_iC_j[y]$

Its formal expression is $W_i[x_m] \ldots W_j[x_{m+1}] \ldots W_j[y_n] \ldots R_i[y_n]$, similar to (21).
    - $W_iR_jC_j[x] - W_jR_iC_j[y]$

Its formal expression is $W_i[x_m] \ldots R_j[x_m] \ldots W_j[y_n] \ldots R_i[y_n]$, similar to (22).

> $R_iW_jC_j[x] - W_jR_iC_j[y]$,

Its formal expression is $R_i[x_m] \ldots W_j[x_{m+1}] \ldots W_j[y_n] \ldots R_i[y_n]$, similar to (23).

6. Based on the definition of POP, if $RW = \{R_iW_j, R_iW_jC_j, R_iW_jC_i, R_iW_jA_i\}$, then the possible PCG can be as follows:

   1) If $p_{ji}[y] = R_jW_i$, it can be in a PCG by:

   > $W_iW_j[x] - R_jW_i[y]$

   Its formal expression is $W_i[x_m] \ldots W_j[x_{m+1}] \ldots R_j[y_n] \ldots W_i[y_{n+1}]$, i.e., Read-Write Skew2. (24)

   > $W_iR_j[x] - R_jW_i[y]$

   Its formal expression is $W_i[x_m] \ldots R_j[x_m] \ldots R_j[y_n] \ldots W_i[y_{n+1}]$, i.e., Read Skew2. (25)

   > $R_iW_j[x] - R_jW_i[y]$

   Its formal expression is $R_i[x_m] \ldots W_j[x_{m+1}] \ldots R_j[y_n] \ldots W_i[y_{n+1}]$, i.e., Write Skew. (26)

   2) If $p_{ji}[y] = R_jW_iC_i$, then $p_{ij}[x]$ must exists $C_i$, therefor exists a PCG

   > $W_iW_jC_i[x] - R_jW_iC_i[y]$, i.e., $p_{ij}[x] = W_iW_jC_i$, therefor exists a PCG, similar to (1).

   > $W_iR_jC_i[x] - R_jW_iC_i[y]$

   Its formal expression is $W_i[x_m] \ldots R_j[x_m] \ldots R_j[y_n] \ldots W_i[y_{n+1}]$, similar to (25).

   > $R_iW_jC_i[x] - R_jW_iC_i[y]$, i.e., $p_{ij}[x] = R_iW_jC_i$,

   Its formal expression is $R_i[x_m] \ldots W_j[x_{m+1}] \ldots R_j[y_n] \ldots W_i[y_{n+1}]$, similar to (26). (26)

   3) If $p_{ji}[y] = R_jW_iC_j$, then $p_{ij}[x]$ must exists $C_j$, therefor exists a PCG

   > $W_iW_jC_j[x] - R_jW_iC_j[y]$,

   Its formal expression is $W_i[x_m] \ldots W_j[x_m] \ldots R_j[y_n] \ldots W_i[y_{n+1}]$, similar to (24).

   > $W_iR_jC_j[x] - R_jW_iC_j[y]$

   Its formal expression is $W_i[x_m] \ldots R_j[x_m] \ldots R_j[y_n] \ldots W_i[y_{n+1}]$, similar to (25).

   > $R_iW_jC_j[x] - R_jW_iC_j[y]$, i.e., $p_{ij}[x] = W_jW_iC_j$,

   Its formal expression is $R_i[x_m] \ldots W_j[x_m] \ldots R_j[y_n] \ldots W_i[y_{n+1}]$, similar to (26).

   4) If $p_{ji}[y] = R_jW_iA_j$, then $p_{ij}[x]$ must exists $A_j$, therefor exists a PCG

   > $W_iW_jA_j[x] - R_jW_iA_j[y]$,

   Its formal expression is $W_i[x_m] \ldots R_j[x_m] \ldots R_j[y_n] \ldots W_i[y_{n+1}]$, similar to (24).

   > $W_iR_jA_j[x] - R_jW_iA_j[y]$, in this case, $t_j$ rollback two read operation, no anomaly occurs.

   Its formal expression is $W_i[x_m] \ldots R_j[x_m] \ldots R_j[y_n] \ldots W_i[y_{n+1}]$, similar to (25).

   > $R_iW_jA_j[x] - R_jW_iA_j[y]$, now, $p_{ij}[x] = W_jW_iC_j$,

   Its formal expression is $R_i[x_m] \ldots W_j[x_m] \ldots R_j[y_n] \ldots W_i[y_{n+1}]$, similar to (26).